\pdfoutput=1
\documentclass[sn-basic]{sn-jnl}
\usepackage{graphicx}%
\usepackage{multirow}%
\usepackage{amsmath,amssymb,amsfonts}%
\usepackage{amsthm}%
\usepackage{mathrsfs}%
\usepackage[title]{appendix}%
\usepackage[table]{xcolor}
\usepackage{textcomp}%
\usepackage{manyfoot}%
\usepackage{booktabs}%
\usepackage{algorithm}%
\usepackage{algorithmicx}%
\usepackage{algpseudocode}%
\usepackage{listings}%
\usepackage{subcaption}
\usepackage{rotating}
\usepackage{diagbox}
\usepackage{verbatim}
\usepackage{tikz}
\usetikzlibrary{shapes,arrows,positioning}
\usepackage[T1]{fontenc}
\begin{document}
	
	\title{A Survey on Point-of-Interest Recommendations Leveraging Heterogeneous Data}
	
	\author[1,2]{\fnm{Zehui} \sur{Wang}}\email{Zehui.Wang@rwu.de}
	
	\author[1]{\fnm{Wolfram} \sur{Höpken}}\email{Wolfram.Hoepken@rwu.de}
	
	\author*[2]{\fnm{Dietmar} \sur{Jannach}}\email{Dietmar.Jannach@aau.at}
	
	\affil[1]{\orgdiv{Institute for Digital Transformation}, \orgname{University of Applied Sciences Ravensburg-Weingarten}, \orgaddress{\street{Doggenriedstrasse}, \city{Weingarten}, \postcode{88250}, \state{Baden-Württemberg}, \country{Germany}}}
	
	\affil[2]{\orgdiv{Department of Artificial Intelligence and Cybersecurity}, \orgname{University of Klagenfurt}, \orgaddress{\street{Universitätsstraße 65-67}, \city{Klagenfurt am Wörthersee}, \postcode{9020}, \state{Kärnten}, \country{Austria}}}

	\abstract{Tourism is an important application domain for recommender systems. In this domain, recommender systems are for example tasked with providing personalized recommendations for transportation, accommodation, points-of-interest (POIs), etc. Among these tasks, in particular the problem of recommending POIs that are of likely interest to individual tourists has gained growing attention in recent years. Providing POI recommendations to tourists can however be especially challenging due to the variability of the user's context. With the rapid development of the Web and today's multitude of online services, vast amounts of data from various sources have become available, and these heterogeneous data represent a huge potential to better address the challenges of POI recommendation problems. In this work, we provide a survey of published research on the problem of POI recommendation between 2021 and 2023. The literature was surveyed to identify the information types, techniques and evaluation methods employed. Based on the analysis, it was observed that the current research tends to focus on a relatively narrow range of information types and there is a significant potential in improving POI recommendation by leveraging heterogeneous data. As the first information-centric survey on POI recommendation research, this study serves as a reference for researchers aiming to develop increasingly accurate, personalized and context-aware POI recommender systems.
	}

	\keywords{Recommender Systems, Tourism, Point-of-Interest Recommendation, Heterogeneous Data}
	
	\maketitle

	\section{Introduction}\label{Introduction}
\begin{figure}[!htb]%
		\centering
		\includegraphics[width=1\textwidth]{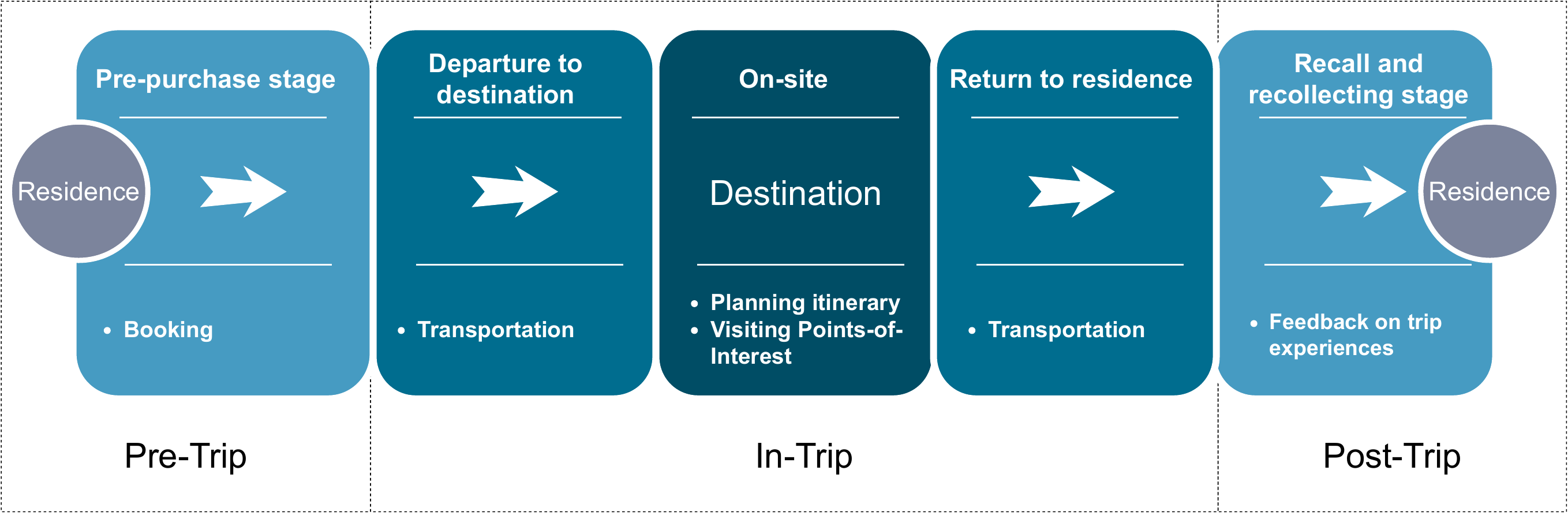}
		\caption{Temporal Phases of the Tourism Process}\label{trip}
	\end{figure}

	Tourism is the act of traveling for pleasure or business to places outside one's usual environment \citep{HAMID2021100337}. It includes a wide range of activities such as visiting tourist attractions, sightseeing, participating in cultural activities and exploring natural wonders. Clawson and Knetsch were the first to conceptualize recreational experiences as a multiphase process~\citep{five_phase_tourism}, consisting of five  parts: (1) anticipation or pre-purchase, (2) the travel to the destination segment, (3) the on-site experience, (4) the return travel segment and (5) an extended recall and recollection stage~\citep{clawson2013economics}. As a common alternative to Clawson and Knetsch's five-phase model, a tourism process can typically be divided into three distinct phases along the temporal dimension \citep{three_phase}: pre-trip, in-trip and post-trip, as illustrated in Figure~\ref{trip}.
	
	The pre-trip and in-trip stages are the periods during which tourists make decisions, such as selecting destinations, choosing transportation modes, and adjusting their visiting time or tourist attractions based on the changing contextual environment \citep{TripDecision}.
	Among the various aspects of trip planning, selecting appropriate POIs can be a significant challenge for tourists, since \emph{point-of-interest} is a holistic concept that encompasses any places tourists can visit during their trip, including museums, parks, cinemas, art galleries, restaurants, coffee shops, shopping centers, etc.~\citep{electronics11131998}. And it can be a time-consuming task for tourists to filter out relevant content from the vast amount of available information about POIs on the Internet.
	
	In order to address these issues, information mechanisms are urgently needed in the tourism domain to assist users by making useful and effective suggestions from the plethora of available POI choices. Recommender Systems (RSs) are considered as established solutions for this task due to their ability to provide personalized recommendations based on various travel purposes and individual preferences \citep{Ricci2022}. Offering POI recommendations may however face significant challenges in providing up-to-date recommendations based on tourists’ preferences and context \citep{graphIntegration}. To achieve this, POI RSs require access to user-related data to understand their needs, preferences and the current context. Therefore, it is crucial to collect and analyze all kinds of available data in the tourism domain to offer valuable recommendations to visit POIs.
	
	Tourists engage in various activities and interact with various elements throughout the duration of their trips. With the widespread use of smartphones and various online applications, a wealth of tourism-related information can be obtained from multiple data sources. These sources encompass descriptions and statistics related to tourism offers and marketing, as well as records of tourists' feedback on consumption of tourism products and services \citep{80}. Such information constitutes a fertile ground for investigating tourists' preferences and behavior in the context of POI recommendation research. However, integrating the aforementioned information types to address the recommendation challenges during trips poses significant difficulties due to their inherent heterogeneity, which manifests in a high variability in data types and formats  \citep{wang2017heterogeneous}.
	
	These different information types can exhibit heterogeneity from syntactic, conceptual, terminological, semiotic and other aspects due to the diverse demographic backgrounds (language, age, gender, etc.) of the data generators, the variety of data acquisition devices (mobile phones, computers, GPS devices, etc.) and the complexity of data types (text, images, videos, trajectories, etc.) \citep{JirkovskyO14}. This issue also exists in the field of RS for tourism, since integrating different information types to establish a comprehensive user model is a challenge. Therefore, data integration techniques that can effectively analyze and explore comprehensive data have become more important in recent years \citep{MeriemeMCFG22}. However, it is important to note that despite the potential benefits, the integration of heterogeneous data as input to POI RSs lacks a systematic and comprehensive overview of its utilization in the existing literature.
	Although extensive research exists on tourism POI recommendation, with recent advancements in multimodal models, to the best of our knowledge no survey in the past three years has reviewed the use of heterogeneous data and multimodal models in tourism POI recommendation.
	To bridge this gap, the contribution of this work is as follows:
	
	\begin{itemize}
		\item We conduct a systematic literature review spanning from January 2021 to September 2023. The review provides an analysis of widely used information types, state-of-the-art techniques and popular evaluation metrics in the context of POI recommendations.
		
		\item We review the current utilization of heterogeneous data in the field of POI recommendation research. Our study offers insights into the heterogeneous types of data that have been employed and sheds light on the integration techniques utilized in existing POI RSs.
		
		\item We identify and present potential research directions that can guide future work in the field of POI recommendation. These research directions provide insights and serve as a roadmap for researchers to explore novel approaches, address existing challenges and advance the state-of-the-art POI RSs.
	\end{itemize}
	
	The rest of the paper is structured as follows: Section \ref{Background and Preliminaries} provides  background information and preliminaries about RSs in the POI recommendation domain; Section \ref{Research Methodology} describes the research methodology of our information-centric survey for POI recommendations; Section \ref{A Landscape of Research} presents the main findings derived from collected papers; Section \ref{Discussion} discusses the current state of research on the integration of heterogeneous data and outlines open gaps and future research opportunities; Section \ref{Conclusion} summarizes the analysis from this work in response to the proposed research questions.
	
	\section{Background and Preliminaries}\label{Background and Preliminaries}
	In this section, we present an overview of the fundamental concepts associated with RSs and their specific application within the tourism domain. Furthermore, we provide a concise summary of past surveys conducted in the realm of POI recommendation, elucidating the gaps and constraints in the existing survey literature.
	
	\subsection{Recommender Systems and Their Applications in Tourism}
	RSs have become ubiquitous in various application domains today, and their origins can be traced back to the early 1990s, when they were first applied in experimental settings for personal email and information filtering \citep{JannachPRZ21}. Since then, they have become a common feature of many online platforms, serving as tools for helping users discover content that may be of interest to them. With the continual progress and evolution of recommender systems, an array of diverse system types has emerged.

	\subsubsection{Types of Recommender Systems}\label{Types of Recommender Systems}
	The most utilized types of recommender systems comprise collaborative filtering (CF) methods, content-based (CB) methods and  hybrid methods \citep{Ricci2022}.
	\paragraph{CF methods}
	CF methods are a class of recommendation systems that utilize explicit or implicit interactions between users and items to provide personalized recommendations. The basic idea of collaborative recommendation approaches is to leverage the historical behavior and opinions of an existing user community to predict which items the current user is likely to prefer or find interesting. Based on how user-item interactions are processed, CF methods can be further categorized into two primary techniques: memory-based and model-based.
	
	\begin{itemize}
		\item \textbf{Memory-based CF methods}:
		In memory-based CF RSs, recommendations are generated by directly comparing user-item interactions. This approach involves computing either user similarities (user-based) or item similarities (item-based). User-based CF suggests items to a target user based on the preferences of users with similar behavior, while item-based CF recommends items based on their similarity to items the user has previously shown interest in. Memory-based CF methods are theoretically more precise because the complete interaction data is available for generating recommendations. However, these systems encounter scalability challenges when dealing with databases containing  millions of users and  items \citep{RSdefinition}.
		
		\item \textbf{Model-based CF methods}:
		In contrast, model-based CF RSs utilize various techniques, such as matrix factorization, association rule mining, probabilistic approaches or deep learning (DL) models, to process raw interaction data and make recommendations through precomputed models. The advantage of model-based CF methods lies in its ability to handle sparse data and scale to larger datasets. However, it may demand more computational resources and expertise for development and deployment \citep{ModelBased}.
	\end{itemize}
	
	While pure CF methods do not necessitate knowledge about the items themselves, they do face challenges when it comes to handling new users or items with limited interaction history, commonly referred to as the cold start problem. Additionally, these methods may struggle when recommending less popular items or accommodating niche user preferences, which is known as the data sparsity problem. Furthermore, privacy concerns can arise due to the need for user behavior data. Hence, it is crucial to investigate alternative methods  that can effectively overcome these constraints and offer personalized recommendations.
	
	\paragraph{CB methods}
	CB methods represent a class of recommendation systems that  primarily depend on the inherent characteristics and features of items to offer personalized recommendations.  The fundamental principle of CB methods is to analyze and understand characteristics and features of an item as \emph{content}. Subsequently, recommendations are made based on these item content to suggest items that are similar to those the user liked in the past.
	
	When compared to CF-based methods, CB methods do not require large user datasets to achieve reasonable recommendation accuracy. These methods can promptly recommend new items as soon as their content becomes available \citep{RSdefinition}. Nevertheless, pure content-based recommender systems are associated with well-known limitations \citep{CBLimitation}, such as shallow item content analysis, which can result in recommendations that do not align closely with user preferences. CB methods can also exhibit item overspecialization, leading to a lack of serendipity. These identified limitations have catalyzed the development of hybrid systems that amalgamate the strengths of different recommendation techniques.
	
	\paragraph{Hybrid methods}
	Hybrid methods represent a class of RSs that can leverage multiple types of data and can be based on multiple algorithmic techniques, as well as demographic-based and knowledge-based RS approaches~\citep{Burke2007}. Hybrid RSs have been devised to overcome the limitations of individual methods, especially in scenarios with access to multiple data sources. They aim to amalgamate insights from diverse data sources and utilize the algorithmic strengths of various RSs for more robust and effective recommendations.  More detailed insights into these combinations can be found in \cite{hybridRS}.
	
	\subsubsection{Applications of Recommender Systems in Tourism}
	With the help of the aforementioned techniques, RSs have been extensively applied in various domains such as e-commerce \citep{Schafer1999RS,ZankerBricmanEtAl2006}, social media \citep{9739361}, entertainment \citep{Schedl2022,s22134904}, and education \citep{3278071}. As an important application, RSs in the tourism domain can recommend tourists the most appropriate transportation options (such as flights and trains), accommodations, POIs and other items that are necessary for their trip \citep{sarkar2023tourism}. Therefore, in the present era of information overload, there is an increasing demand for Tourism Recommender Systems (TRSs) to reduce the time spent on information retrieval and decision making for travel. These systems are designed to not only reduce the time spent on retrieving travel information but also to provide several additional advantages to users:
	
	\begin{itemize}
		\item \textbf{Personalization}: TRSs offer a high degree of personalization in travel recommendations, tailoring suggestions to individual preferences and leveraging historical user behavior data. This not only enhances the overall travel experience but also fosters user engagement and satisfaction~\citep{BORRAS20147370}.
		
		\item \textbf{Broad Applicability}: TRSs provide users with a comprehensive array of travel-related recommendations, encompassing various aspects of the travel experience such as accommodations, attractions, dining options, and leisure activities. This diversity ensures that users receive a holistic and enriched travel itinerary~\citep{Chaudhari2020}.
		
		\item \textbf{Local Insights}: TRSs often incorporate localized knowledge and insights, providing users with the opportunity to discover less-explored attractions and engage with destinations from an authentic local perspective. This feature fosters a deeper and more culturally immersive travel experience~\citep{MISHRA2023100145}.
		
	\end{itemize}
	
\begin{figure}[!htb]
		\centering
		\resizebox{\textwidth}{!}{
			\begin{tikzpicture}[node distance=1cm]
				
				\node [rectangle, draw, thick, text width=5cm, align=center, font=\large] (trs) {\textbf{Tourism  \\ Recommender Systems}};
				
				\node [rectangle, draw, thick, text width=3.5cm, align=center, below=of trs, xshift=-9cm] (accommodation) {\textbf{Accommodation \\ Recommendation}};
				\node [rectangle, draw, thick, text width=3.5cm, align=center, below=of trs, xshift=-4cm] (dining) {\textbf{Dining \\ Recommendation}};
				\node [rectangle, draw, thick, text width=3.5cm, align=center, below=of trs, xshift=2.5cm] (destination) {\textbf{Attraction \\  Recommendation}};
				\node [rectangle, draw, thick, text width=3.5cm, align=center, below=of trs, xshift=9cm] (other) {\textbf{Other \\ Recommendation}};
				
				\node [rectangle, draw, text width=2.5cm, align=center, below =of accommodation, yshift=-0.2cm] (hotel) {\textbf{Hotel} \\ e.g., \cite{hotel}};
				
				\node [rectangle, draw, text width=2.5cm, align=center, below left=of dining, xshift=2cm, yshift=-0.5cm] (restaurant) {\textbf{Restaurant} \\ e.g., \cite{restaurant}};
				
				\node [rectangle, draw, text width=3.5cm, align=center, below right=of dining, xshift=-3cm, yshift=-0.5cm] (food) {\textbf{Food} \\ e.g., \cite{food}};
				
				\node [rectangle, draw, text width=2cm, align=center, below left=of destination, xshift=2cm, yshift=-0.5cm] (itinerary) {\textbf{Itinerary} \\ e.g., \cite{itineray}};
				\node [rectangle, draw, text width=2cm, align=center, below left=of destination, xshift=5cm, yshift=-0.5cm] (poi) {\textbf{Point-of-Interest}};
				
				\node [rectangle, draw, text width=2.5cm, align=center, below left=of poi, xshift= 1.5cm, yshift=-1cm] (general) {\textbf{General POI}};
				\node [rectangle, draw, text width=2cm, align=center, below right=of poi, xshift= -1.5cm, yshift=-1cm] (next) {\textbf{Next-POI}};
				
				\node [rectangle, draw, text width=3cm, align=center, below=of other, xshift= -2cm, yshift=-0.2cm] (transportation) {\textbf{Transportation} \\ e.g., \cite{transportation}};
				\node [rectangle, draw, text width=3.5cm, align=center, below=of other, xshift= 0cm, yshift=-2.3cm] (photo) {\textbf{Photography} \\ e.g., \cite{photo}};
				\node [rectangle, draw, text width=2cm, align=center, below=of other, xshift= 2cm, yshift=-0.2cm] (outfit) {\textbf{Outfits} \\ e.g., \cite{outfit}};
				
				\draw[-, line width=1pt] (trs) -- ++(-3,0) -| (accommodation);
				\draw[-, line width=1pt] (trs) -- (dining);
				\draw[-, line width=2pt] (trs) -- (destination);
				\draw[-, line width=1pt] (trs) -- ++(3,0) -| (other);
				
				\draw[-, line width=1pt] (accommodation) -- (hotel);
				
				\draw[-, line width=1pt] (dining) -- (restaurant);
				\draw[-, line width=1pt] (dining) -- (food);
				
				\draw[-, line width=1pt] (destination) -- (itinerary);
				\draw[-, line width=2pt] (destination) -- (poi);
				
				\draw[->, line width=2pt] (poi) -- (general);
				\draw[->, line width=2pt] (poi) -- (next);
				
				\draw[-, line width=1pt] (other) -- (transportation);
				\draw[-, line width=1pt] (other) -- (photo);
				\draw[-, line width=1pt] (other) -- (outfit);
				
			\end{tikzpicture}
		}
		\caption{Main Recommendation Tasks in Tourism}\label{TRS}
	\end{figure}
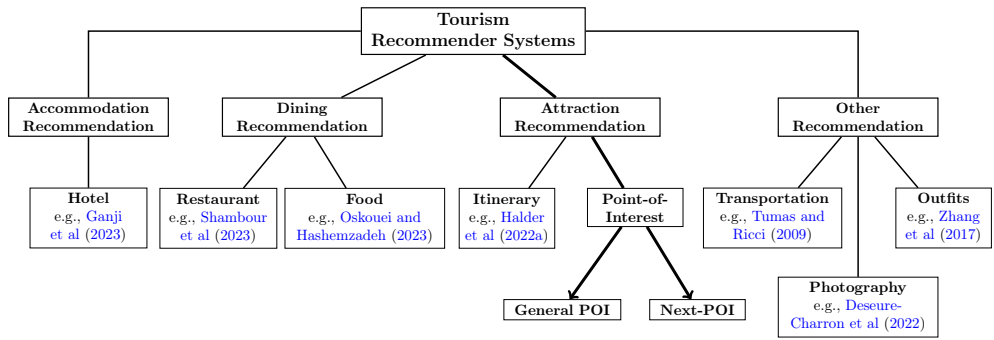

	By combining these advantages, TRSs play a pivotal role in simplifying and enhancing the travel planning process. They effectively assist users with a variety of tourism-related recommendations \citep{Chaudhari2020}. Based on the findings from \cite{Chaudhari2020}, we have listed the main application domains of TRS in Figure~\ref{TRS}. To provide readers with a clearer understanding of TRS applications in each domain, Figure~\ref{TRS} also lists examples of pertinent works.
	
	When examining the application of TRSs specifically in the domain of POI recommendation\footnote{In this survey, POI recommendation refers exclusively to \textit{tourism-related} POI recommendation, meaning the recommendation of any locations that tourists may visit during their trip.
	}, it becomes apparent that data can be gathered throughout various stages of the trip, including tourism business transactions. Examples of such data are presented in Table~\ref{BI} reflecting tourists' preferences for POIs and providing valuable information for building effective POI RSs, as discussed in \citep{Höpken2022}. Aside from these information collected from tourism business transactions at each stage of the trip, there are other information types that can be utilized to generate personalized recommendations. These include tourists’ demographic information and friendships from profiles on social media platforms \citep{KolahkajHNC20,12}, as well as basic information about tourism products and services, such as POI location, costs, facilities \citep{QomariyahK21}, etc. Moreover, contextual information about traffic and weather conditions during travel can be explored to build context-aware RS as well \citep{ZhuWCSYC18,7}. By leveraging these diverse information types, POI RSs can generate personalized recommendations that align with the individual preferences and needs of tourists throughout their entire trips. The integration and analysis of these data from different sources offer the potential to enhance the accuracy and relevance of POI recommendation, ultimately enriching the overall travel experience for tourists.
	
	\begin{table}[!htb]
		\centering
		\centering
		\caption{Data Sources for POI Recommendations at Different Trip Phases from Tourists' Business Transaction \citep{Höpken2022}}\label{BI}
		\begin{tabular}{cp{5cm}p{5cm}}
			\toprule
			\textbf{Phase}       & \multicolumn{1}{c}{\textbf{Data}}   & \multicolumn{1}{c}{\textbf{Source}}   \\ \hline
			& Descriptions and marketing statistics of tourism products or services  & Official websites, marketing networks, search engines, social media platforms                \\ \cmidrule{2-3}
			Pre-trip             &\multirow{2}{*}{Information search records}   & Search engines, travel websites, mobile apps/guides, social media platforms\\ \cmidrule{2-3}
			\multicolumn{1}{l}{} & Booking or reservation records    & Booking systems \\ \hline
			& Transportation trajectories and POI check-in records & \multirow{2}{*}{Ticket systems, social media platforms}                                              \\ \cmidrule{2-3}
			In-trip          & \multirow{2}{*}{Accommodation records}  & Accommodation providers, official statistics, mobile app usage  \\ \cmidrule{2-3}
			\multicolumn{1}{l}{} & \multirow{2}{*}{Consumption records}   & Local ticket offices, tourists' payment systems  \\ \hline
			\multirow{3}{*}{Post-trip}            & \multirow{3}{*}{Feedback for tourism products}  & Online review sites, supplier-specific online feedback, survey systems, social media platforms \\ \bottomrule
		\end{tabular}
	\end{table}
	
	\subsection{Problem Formulation}\label{Problem Formulation}
	In the context of this work, we formally define the problem of POI recommendation as follows. In the POI recommendation scenario, given a POI visiting history of a user (including POI identifier, time and location) and additional contextual information, our objective is to generate personalized recommendations for this user.
	Let  $\mathcal{U}=\{ u_1, \dots, u_ { \left | \mathcal{U} \right |} \}$ denote a set of users, $\mathcal{P}=\{ p_1, \dots, p_ { \left | \mathcal{P} \right |} \}$ denote a set of POIs\footnote{POI $p$ refers solely to the POI ID, enabling a clearer exploration of the influence of heterogeneous information on POI recommendations \citep{34,38}.}, $\mathcal{Q}^{u}=\{ q_1^{t_1}, \dots, q_{ \left | \mathcal{Q} \right |}^{t_{\left | \mathcal{T} \right |}} \}$ denote a visit event sequence\footnote{In the existing literature, these events are often captured through check-in events.} of user $u$, and  $\mathcal{T}=\{ t_1, \dots, t_ { \left | \mathcal{T} \right |} \}$ present a set of time periods. Each POI visitation, i.e., check-in, is denoted by $q =< u,p,t,g,c>$, expressing that user $u$ visited a POI  $p$ at timestamp $t$ under certain contextual factors $c$. The geographical information of  POI $p$ is $g=< latitude, longitude>$, and the contextual information $c$ includes additional factors such as weather, traffic conditions to the visit, or tourist’s social relationships. Two types of POI recommendation problems can then be formulated as follows \citep{27-m,25-m}:
	
	\textbf{General POI Recommendation:} For a given target user $u$ and his visited POI list extracted from $\mathcal{Q}^u$, the goal is to generate a top-k list of unvisited POIs within the search space of POIs from set $\mathcal{P}$ that user $u$ is most likely to visit.
	
	\textbf{Next-POI Recommendation:} For a given target user $u$ and his visited POI list extracted from $\mathcal{Q}^u$ during timestamp $t_1$ to $t_m$, the goal is to produce a top-k list of unvisited POIs within the search space of POIs from set $\mathcal{P}$ that user $u$ is most likely to visit at the next timestamp $ t_ {m+1}$.
	
	The primary distinction lies in the fact that in general POI recommendations, the sequence of events does not necessarily influence the recommendations. In contrast, the objective of next-POI recommendations is to utilize sequential patterns to deliver recommendations that depend on a user's most recently visited POIs. It is worth noting that in practical approaches to address both general and next-POI recommendation problems, various types of side information beyond temporal and spatial details are frequently leveraged. These may include factors such as the category of the POI, the social relationships of tourists and the context of the trip.
	If the data used in the POI recommendation process comes from a single type, such as check-in records or user rating scores, it is defined as a \textit{POI recommendation with homogeneous data} in this paper. In contrast, if the data comes from multiple types, combining different information modalities such as social relationships, review texts, images, weather conditions, or traffic data, it is defined as a \textit{POI recommendation with heterogeneous data}.
	The utilization of multiple types of information will be further discussed in Section \ref{Discussion}.
	
	\subsection{Previous Reviews on Point-of-Interest Recommendation}
	With an increasing emergence of research on POI RSs in recent years, a plethora of models have been proposed to tackle the problem of providing personalized POI recommendations. There have been several review articles highlighting the major findings and limitations from different perspectives.
	
	In the light of numerous POI recommendation techniques, a recent study by \cite{134} provided an evaluation of twelve state-of-the-art POI recommendation models. Through this thorough evaluation, significant findings about the different model performances due to data, users or modeling methods were obtained, which can aid in the better understanding and utilization of POI recommendation models in various scenarios.
	Furthermore, \cite{59} conducted a systematic overview of $74$ relevant papers published from 2017 to 2019 and proposed an extensible POI recommendation benchmark to address and identify limitations, including a prioritization of accuracy over other quality dimensions and a low intersection of metrics and datasets used to evaluate proposed solutions. In a subsequent work, \cite{2} developed a reproducibility framework based on Python software libraries and a Docker image to reproduce experimental evaluations on POI recommendations using different datasets, metrics and baselines.
	Due to the surge of research activities utilizing DL paradigms in the field of POI recommendation, a survey of major DL-based POI recommendation works has been compiled by \cite{3}. This survey categorized and critically analyzed recent POI recommendation works based on different DL paradigms, as well as relevant features such as problem formulations, proposed techniques, and used datasets.
	The aforementioned surveys provide valuable insights into current POI recommendation research. Nevertheless, instead of focusing on techniques in POI RSs, our work emphasizes the various information types involved in the POI recommendation process.
	
	From the viewpoint of the data that are used for POI recommendations,
	a comprehensive analysis of the effect of contextual factors, including social, temporal, spatial and categorical factors on recommendation models was conducted by \cite{4},
	using both existing and novel linear/non-linear models.
	Additionally, \cite{80} conducted a survey on the use of Linked Open Data (LOD) in location-based recommendation systems within the tourism domain. This survey provided a systematic review and summarized research achievements between 2001 and 2018. It mapped various aspects of LOD by analyzing how heterogeneous data, such as geographic, statistical, and user-generated content, are integrated and used for tourism-related recommendations. The study also classified the different categories of location-based recommendation applications that use LOD and suggested possible future research directions for enhancing LOD in location-based recommendations for tourism.
	Another survey conducted by \cite{3510409} delved into the domain of POI recommendation research spanning the period from 2011 to 2020, with a specific focus on the data from location-based social networks (LSBNs). The authors provided an intricate analysis of diverse information sources, evaluation methodologies and algorithms within the context of POI recommendation and highlighted both the existing prospects and challenges that continue to persist within this field.
	These aforementioned surveys are closely related with our current study, as they have similarly focused on utilized data in the POI recommendation.
	However, our present survey is not limited to specific data sources, such as data from LBSNs or LOD, but specifically analyzes the utilization of all kinds of information as input to POI RSs based on journal and conference papers published up to September 2023.
	
	In the realm of POI recommendation, despite the growing research interest, there is still a notable absence of a comprehensive, systematic and information-centric comparative analysis that accurately mirrors the state-of-the-art in this field, especially since 2021. In particular, there exists a significant gap in our understanding of how recent research has harnessed heterogeneous data (not only from publicly available datasets but also those collected autonomously, such as through self-crawling or official APIs), and how data integration methods have been employed. In an effort to fill this gap, our study aims to provide an overview, focusing primarily on information types, methodologies and appropriate evaluation metrics.
	
	\section{Research Methodology}\label{Research Methodology}
	This section outlines the research methodology adopted to conduct a systematic literature review and to gather relevant research papers for this study. The primary objective of this research is to offer an overview of the most recent advancements in the realm of POI recommendation, specifically focusing on an information-centric perspective.
	
	\subsection{Definition of Research Questions}
	In order to achieve the stated objective, the following research questions (RQs) were formulated:
	\begin{itemize}
		\item \textbf{RQ1}: What is the current state of research on POI recommendations in terms of data, techniques and evaluation?
		\item \textbf{RQ2}: How are heterogeneous data currently being utilized in the POI recommendation research?
		\item \textbf{RQ3}: What are the existing limitations and potential future directions for research and development of POI RSs?
	\end{itemize}
	
	\subsection{Search Strategy}
	A systematic literature search was conducted in the DBLP database to retrieve English-language journal and conference citations published from January 2021 to September 2023.
	DBLP\footnote{\url{https://dblp.org}} is a widely recognized computer science bibliography website renowned for its extensive coverage of computer science and its associated disciplines. In this paper, we have chosen DBLP as our search platform due to its advantageous features of free access and timely updates.
	The time frame January 2021 to September 2023 was selected to focus on recent research and minimize overlap with previous surveys on tourism RSs. The search strategy involved various search queries related to TRS, with a particular emphasis on terms related to the concept of POIs.
	
	Focusing on the bolded sections of Figure~\ref{TRS}, common prefixes were incorporated, such as the inclusion of terms like ``recommend'' for ``recommender system''  or ``recommendation''. To ensure comprehensive coverage of the search results, synonymous terms for the keywords ``POI''  and ``point-of-interest'' were included. The final search query was as follows:
	\textit{``recommend''} AND (\textit{``point-of-interest''} OR \textit{``POI''} OR \textit{``tour''} OR \textit{``activity''} OR \textit{``attraction''} OR \textit{``location''})
	
	Through this search strategy, our aim was to capture a wide range of relevant literature related to RSs, in the context of POI recommendation.
	
	\subsection{Screening of Papers}
	The inclusion criteria to select relevant papers for this study were defined as follows:
	
	\begin{enumerate}
		\item The paper is written in English.
		\item The publication date of the paper falls between January 2021 and September 2023 (both included).
		\item The paper is (a) published in journals that rank in Q1 according to the Scimago Journal \& Country Rank (SJR)\footnote{\url{https://www.scimagojr.com}} or Web of Science Journals Ranking (JCR)\footnote{\url{https://jcr.clarivate.com}} for the publication year, or (b) published in conference proceedings, which are renowned in the field of recommender system research and hold prominence in the  List of International Academic Conferences and Periodicals recommended by the Chinese Computer Federation (CCF)\footnote{Alternative rankings, such as CORE, could be used.  Since a large portion of the relevant research originates from Chinese institutions (with 76 out of the 111 collected papers having the first author affiliated with a Chinese institution), we chose to use the CCF ranking system (\url{https://www.ccf.org.cn/en/Bulletin/2019-05-13/663884.shtml}). We expect that relying on other ranking systems like CORE would not affect the main observations of our study.}, being classified as either A or B-level conferences.
		\item The paper is an original research article related to the general POI recommendation or the next-POI recommendation.
	\end{enumerate}
	
	Exclusion criteria were also defined to exclude papers that do not meet the specific requirements of this study.   Papers fulfilling the following criteria were excluded:
	
	\begin{enumerate}
		\item Papers that focus on POI recommendations for specific populations with unique needs (such as individuals with autism~\citep{autism}), since we are not focused on rare and unique problem settings that demand highly specific information types and potentially non-generalizable algorithmic approaches.
		\item Papers that delved into itinerary recommendations or group recommendations. These topics are beyond the scope of our research, since they are outside the defined scenarios outlined in Section \ref{Problem Formulation}.
		\item Papers that propose only a theoretical framework without utilizing any dataset for experimentation.
	\end{enumerate}
	
	\subsection{Paper Mapping}
\begin{figure}[!htb]%
		\centering
		\includegraphics[width=0.7\textwidth]{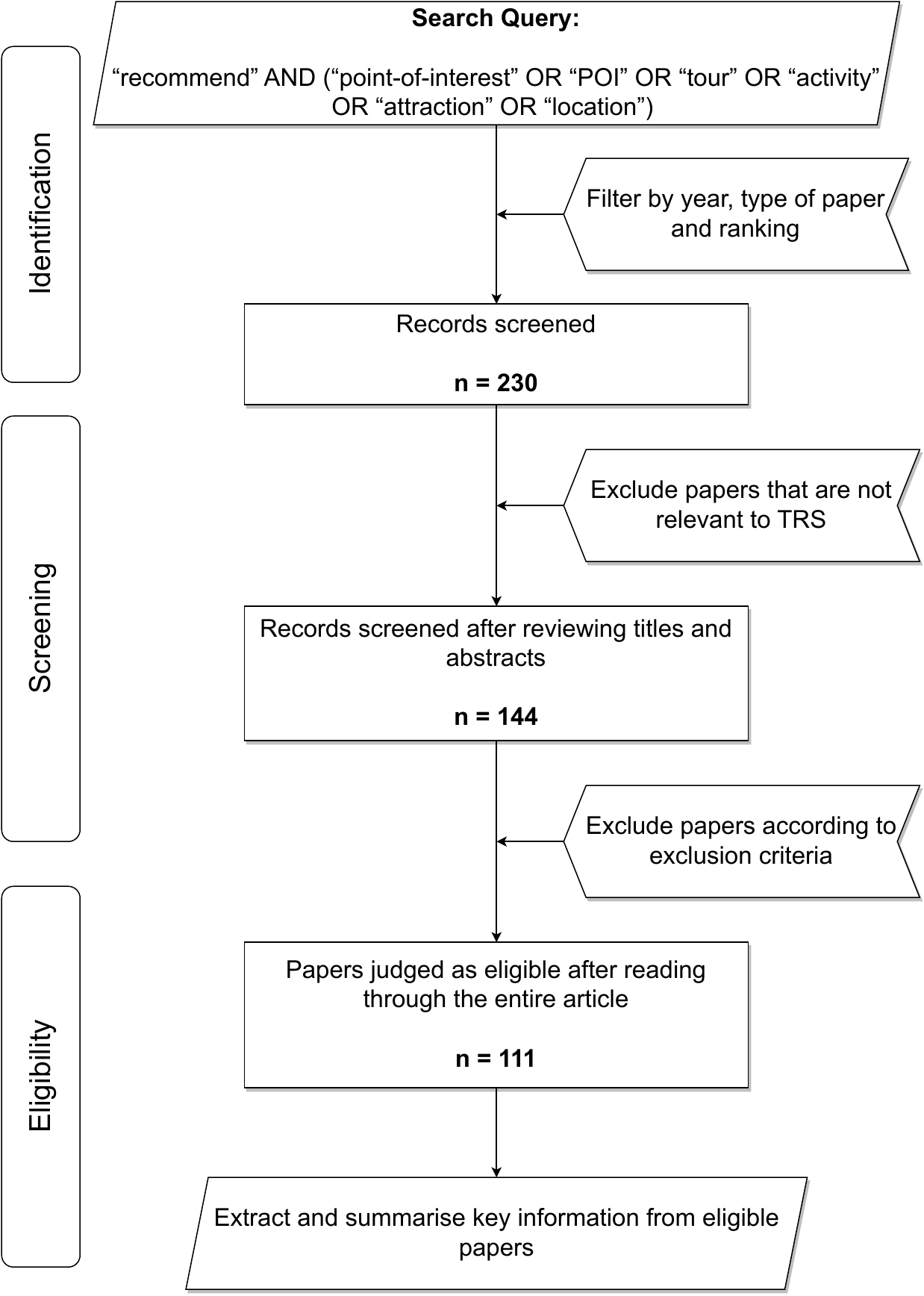}
		\caption{Mapping Process for Identifying Relevant Papers on POI Recommendations}\label{process_diagram}
	\end{figure}

	To identify relevant studies for our review on POI recommendations, we conducted a systematic paper mapping process depicted in Figure~\ref{process_diagram}. The process involved multiple steps to screen, filter and extract key information from the retrieved papers \citep{PAGE2021105906}.
	
	Firstly, we constructed the search query using keywords pertaining to POI recommendations, as outlined above. This query was subsequently employed to query the DBLP database, employing additional filters based on the year of publication, paper type and the ranking of the journal or conference. This process yielded an initial dataset comprising 230 papers.
	
	Next, we conducted a preliminary screening based on the title and abstract of these papers. We reviewed each paper to assess its relevance to the field of TRS and only full-length papers were retained. As a result of this initial screening, along with the elimination of duplicate entries retrieved due to multiple search queries, a total of 144 papers remained in our dataset.
	
	Finally, a further refinement was carried out by excluding papers that met the exclusion criteria outlined above, resulting in a final set of 111 eligible papers. We thoroughly read and analyzed these papers to extract and summarize the key information pertinent to our research objectives.
	
	Overall, the paper mapping process ensured a rigorous and systematic approach in selecting relevant studies for our review. The chosen papers provide valuable insights into the current state-of-the-art POI recommendation research, which will be presented in detail in Section \ref{A Landscape of Research}. The list of papers considered in this review can be found online\footnote{\url{https://github.com/wzehui/Survey_TRS.git}}.
	
	\section{A Landscape of POI Recommendation Research}\label{A Landscape of Research}
	This section provides an overview of the state-of-the-art POI recommendation research from the perspectives of data, techniques and evaluations.
	Specifically, we will first present the results of a statistical analysis of research papers published in journals and conferences from 2021 to 2023, highlighting the trends and patterns in how researchers have approached data collection, technical advancements and evaluation metrics.
	The findings provide a foundation for our further discussion on the research landscape of the use of various information types in the field of POI recommendation.
	
	\subsection{Trends and Developments in POI Recommendations}
\begin{figure}[!htb]%
		\centering
		\begin{subfigure}[b]{0.45\textwidth}
			\centering
			\includegraphics[width=1.2\textwidth]{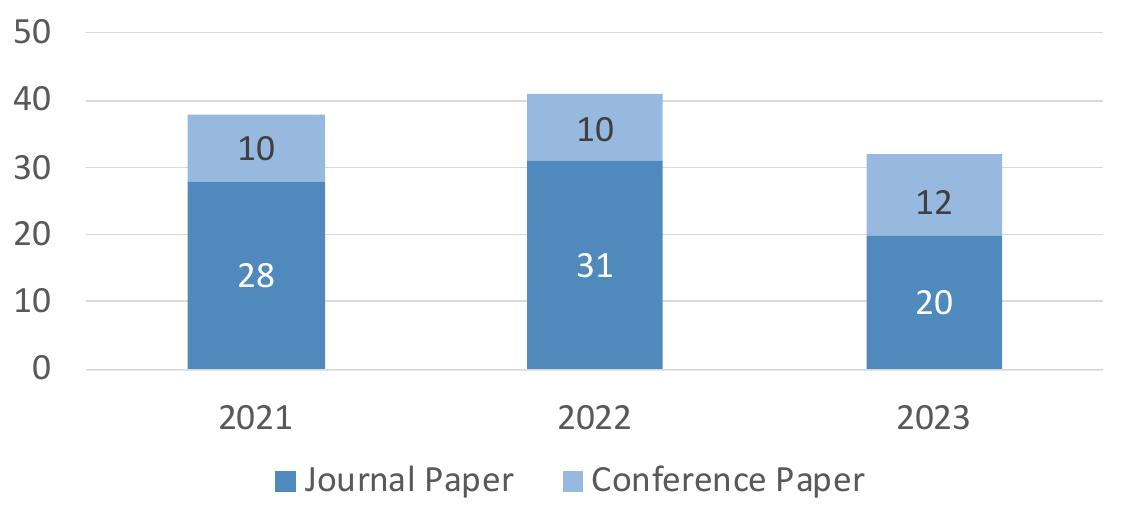}
			\caption{Research Paper Counts by Publication Venue and Year}
			\label{year_distribution}
		\end{subfigure}
		\hfill
		\begin{subfigure}[b]{0.45\textwidth}
			\centering
			\includegraphics[width=1\textwidth]{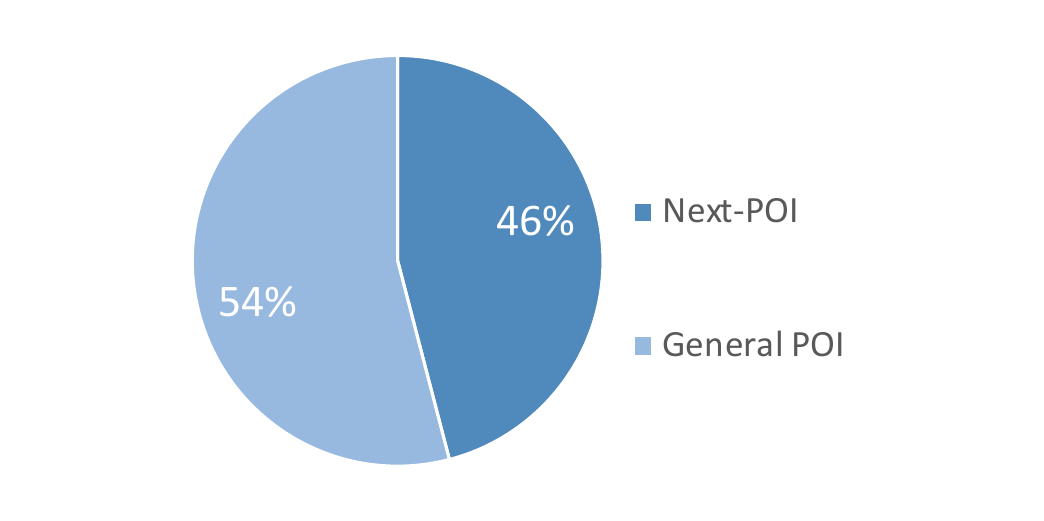}
			\caption{Proportion Distribution of Research Paper Topics}
			\label{domain}
		\end{subfigure}
		\caption{Annual Distribution and Application Focus on POI Recommendations}
	\end{figure}

	Based on the research methodology described in Section \ref{Research Methodology}, a total of $111$ original research papers published between 2021 and 2023 were collected, with their annual distribution depicted in Figure~\ref{year_distribution}.
	Although the papers for 2023 are only available up to September, the results indicate a consistent level of attention to the field of POI recommendation. Based on the findings gathered, the top three journals with the highest number of research papers in this domain are: \emph{Expert Systems With Applications}, \emph{Neurocomputing} and \emph{ACM Transactions on Knowledge Discovery from Data}. Similarly, the top three conferences with the highest paper count are: \emph{SIGIR}, \emph{CIKM}, and \emph{DASFAA}. These findings not only underscore a substantial interest in POI recommendation research but also emphasize the importance of ongoing exploration and evaluation in this field.
	
	As defined in Section \ref{Problem Formulation}, we can broadly categorize the studies into those focusing on the general (sequence-agnostic) POI recommendation or the next-POI recommendation problem. The distribution of research efforts for these topics is illustrated in Figure~\ref{domain}. While the general POI recommendation research maintains its dominant position, the proportion of original research dedicated to the next-POI recommendation has gained increasing attention in the most recent three years. This heightened interest can likely be attributed to the advancements in technologies, such as deep neural networks, enabling a more refined focus on user's short-term preferences. Additionally, sequential recommendation problems in general have also gained a lot of interest in the past few years \citep{sequenceRS}.
	A detailed discussion of this trend will be presented in Section \ref{Data Utilization and Trends in POI Recommendations}. In summary, the heightened attention given to the next-POI recommendation problem in recent research indicates a growing interest for exploring the potential of providing more real-time POI recommendations, which may pave the way for further advancements in this field \citep{RealTime}.
	
	\subsection{Data Utilization in POI Recommendations}\label{Data Utilization and Trends in POI Recommendations}
	As an information-centric survey, we firstly direct our attention towards the status of today's research in terms of data collection and usage in the context of  POI recommendations, since the use of appropriate data plays a crucial role in the development and evaluation of POI RSs.
	
	\subsubsection{Data Collection Approaches and Data Sources}
\begin{figure}[!htb]%
		\centering
		\includegraphics[width=0.8\textwidth]{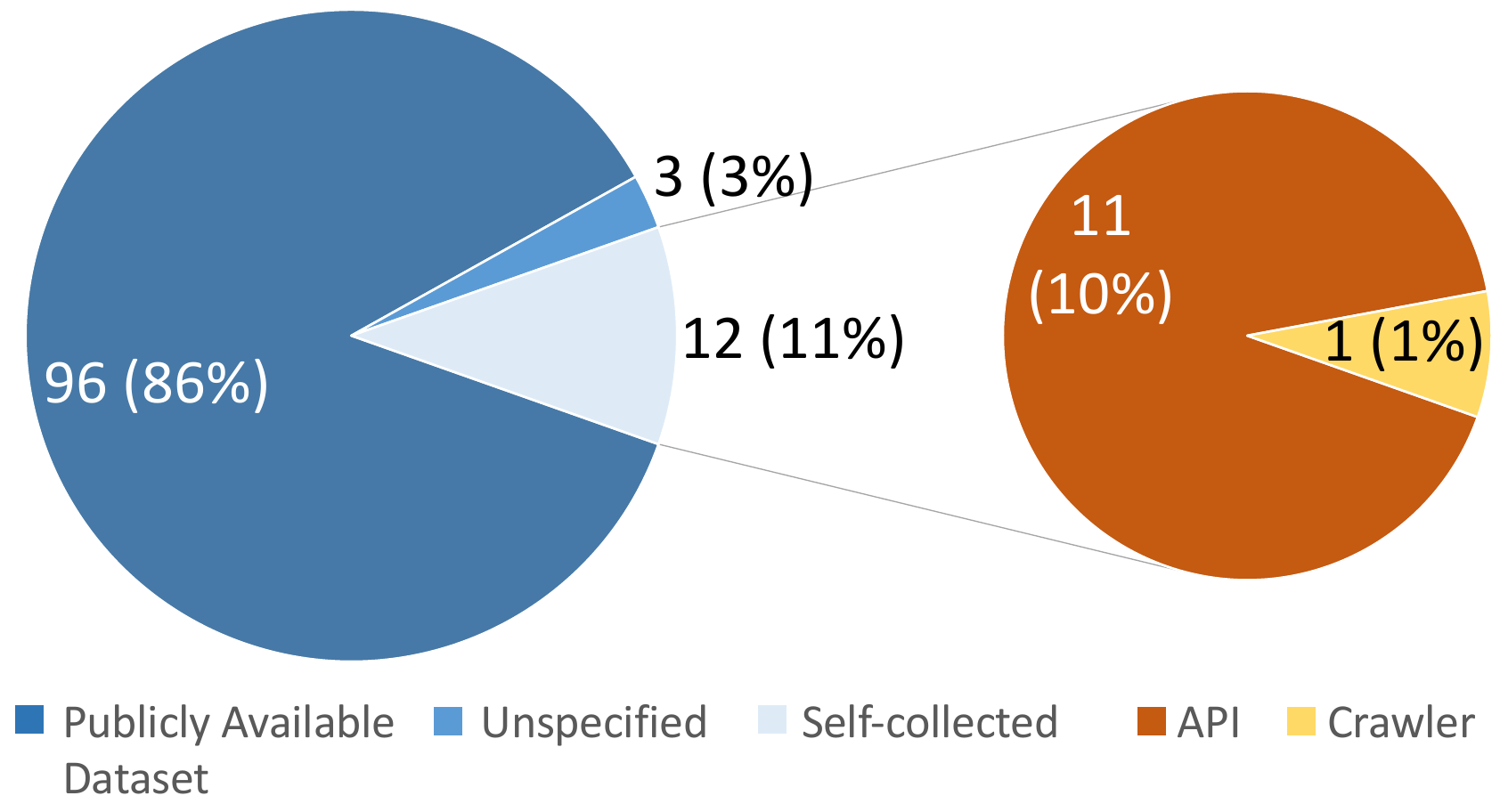}
		\caption{Overview of Data Collection Approaches in POI Recommendation Research}\label{collection_type}
	\end{figure}

	A high-quality dataset can provide valuable insights into user preferences, behavior and context, which are essential for designing effective recommendation algorithms. Figure~\ref{collection_type} provides an overview of different ways for data collection in POI recommendations. We would categorize the primary methods of data collection, as outlined in the gathered papers. For instance, some research incorporates check-in data from publicly available datasets while concurrently gathering contextual information through crawlers or APIs \citep{18-m,13-m}. We categorize papers adopting this approach under the publicly available dataset classification. We can observe that a majority of studies relies on publicly available datasets, which is a readily available and easily accessible source of information. A smaller proportion of studies employs self-collected data either extracted from websites through crawlers or gathered by official APIs. A few papers do not specify how they collect the data in their studies. Overall, the analysis shows  that publicly available datasets are the most frequently used data source in the POI recommendation research due to their convenience and abundance.
	
\begin{figure}[!htb]%
		\centering
		\includegraphics[width=0.6\textwidth]{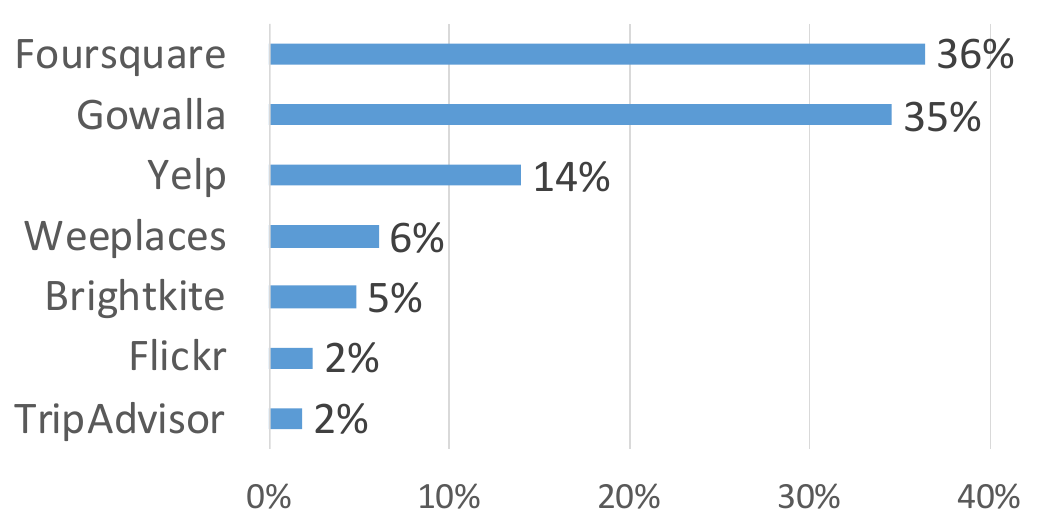}
		\caption{Utilization of Data Source Platforms for Publicly Available Datasets}\label{dataset}
	\end{figure}

	Compared with self-collected data, publicly available datasets typically originate from a transparent and accessible data source. Table~\ref{platform} provides a brief overview of the publicly available datasets frequently used (by at least two studies) in the collected papers, while Figure~\ref{dataset} illustrates their normalized\footnote{For studies that utilized multiple datasets, we aggregate the counts, resulting in normalized percentages for an overview.} utilization proportions. More details about information types included in the datasets mentioned in Table~\ref{platform} can be found in \cite{3}. We do not elaborate further on this in the current paper. Remarkably, the currently available downloadable datasets are primarily sourced from LBSNs. The findings show that more than half of the studies that utilize publicly available datasets make use of data from the Foursquare and Gowalla platforms. In contrast, the utilization of data from other platforms in POI recommendations appears to be relatively limited. It is worth noting that aforementioned platforms offer multiple versions of these datasets, varying in terms of time span, geographical coverage and other relevant characteristics.
	
	\begin{table}[!htb]
		\caption{Overview of Publicly Available Datasets in POI Recommendation Research}\label{platform}
		
		\begin{tabular}{lp{7.5cm}}\toprule
			\textbf{Platform} & \textbf{Description} \\  \midrule
			\emph{Foursquare} & \\ \cmidrule{1-1}
			
			\multirow{3}{*}{Global-scale Check-in Dataset} & 33,278,683 global-scale check-ins by 266,909  users, covering 3,680,126 venues across 415 cities in 77 countries from April 2012 to September 2013; \citep{2814575,YANG2015170} \\ [1cm]
			\multirow{3}{*}{NYC and Tokyo Check-in Dataset} & 227,428 check-ins in New York City and 573,703 check-ins in Tokyo, spanning from 12 April 2012 to 16 February 2013;  \citep{6844862}  \\ [1cm]
			\multirow{2}{*}{Weeplaces Dataset} & 7,658,368 check-ins generated by 15,799 users over 971,309 locations \\ \midrule
			
			\emph{Gowalla} & \\ \cmidrule{1-1}
			\multirow{3}{*}{Stanford Large Network Dataset}  & 6,442,890 check-ins and an undirected friendship network with 196,591 nodes and 950,327 edges between February 2009 and October 2010 \citep{2020579} \\ \midrule
			
			\emph{Yelp} & \\ \cmidrule{1-1}
			\multirow{3}{*}{Yelp Open Dataset} & 908,915 tips provided by 1,987,897 users and aggregated check-in information over time for each of the 131,930 businesses   \\ \midrule
			
			\emph{Brightkite} & \\ \cmidrule{1-1}
			\multirow{3}{*}{Stanford Large Network Dataset} & 4,491,143 check-ins and an undirected friendship network with 58,228 nodes and 214,078 edges between April 2008 and October 2010 \citep{2020579}    \\ \midrule
			\emph{Flickr} \\  \cmidrule{1-1}
			\multirow{2}{*}{YFCC100M} & 99.2 million photos and 0.8 million videos, spanning from 2004 until early 2014 \citep{2812802} \\ [0.7cm]
			\multirow{3}{*}{Flickr User-POI Visits Dataset} & A set of users and their visits to various POIs in 8 cities, which are determined based on YFCC100M Flickr photos \cite{2832496,3038646}     \\ \midrule
			\emph{TripAdvisor} \\ \cmidrule{1-1}
			\multirow{4}{*}{TripAdvisor Dataset} & Ratings for POIs in the South Tyrol region of Italy that are tagged with contextual situations described by the conjunction of contextual conditions coming from type, month and year of the trip \citep{braunhofer2016contextual} \\
			
			\bottomrule
		\end{tabular}
	\end{table}
	
	\subsubsection{Information Type Analysis}
\begin{figure}[!htb]
		\centering
		\includegraphics[width=1\columnwidth]{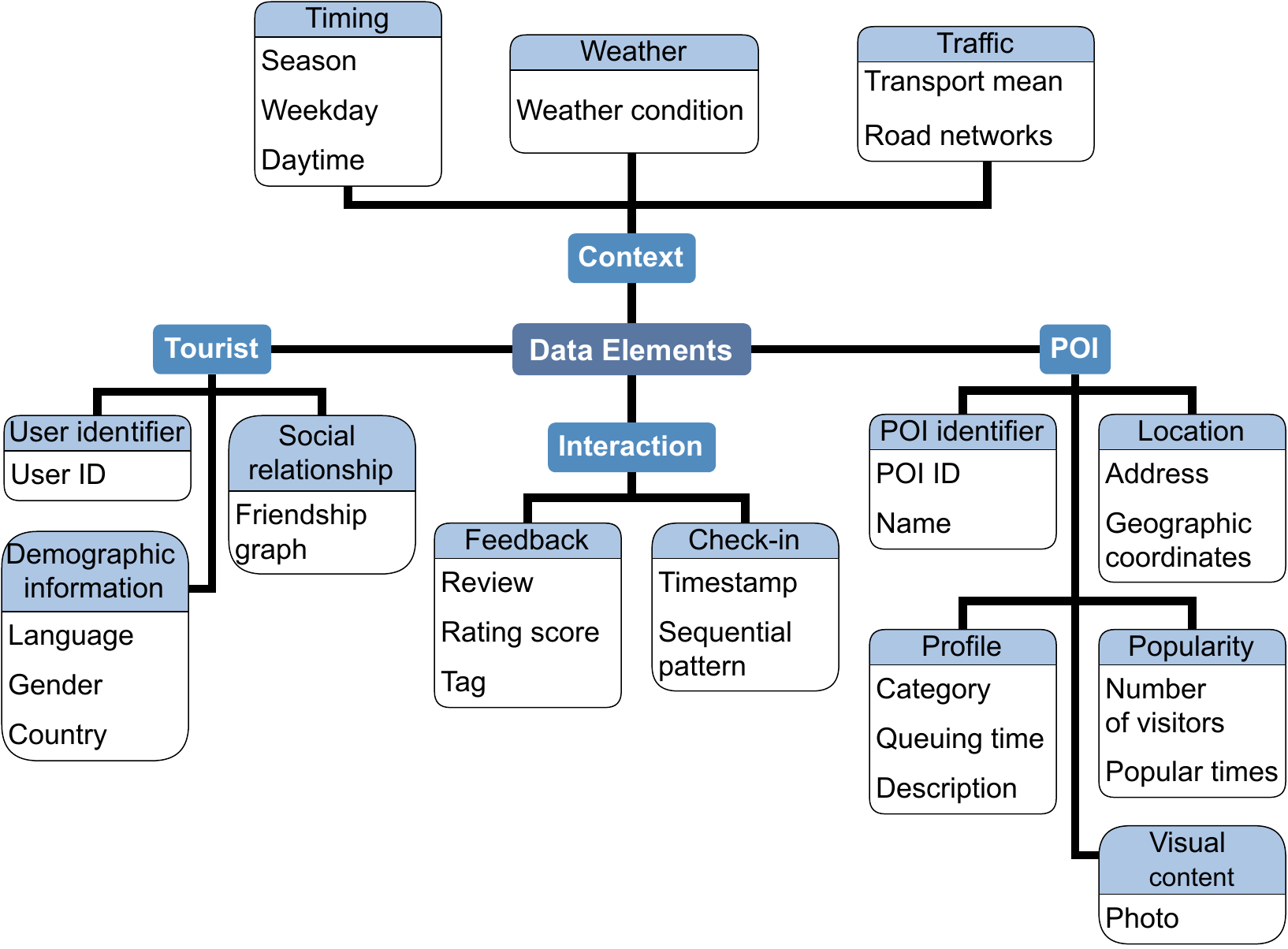}
		\caption{Taxonomy of Data Elements for POI Recommendations}\label{up_buttom}
	\end{figure}

\begin{figure}[!htb]%
		\centering
		\includegraphics[width=0.7\textwidth]{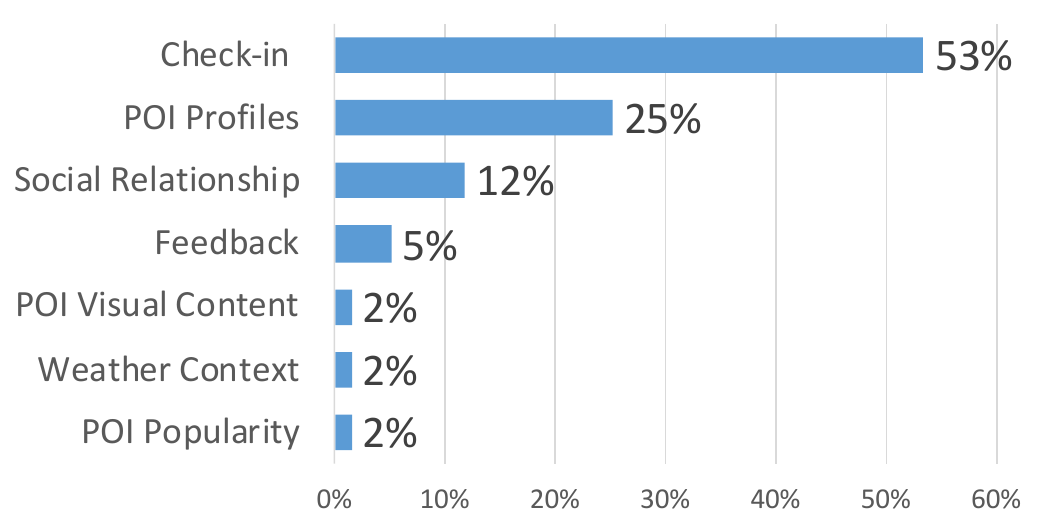}
		\caption{Information Type Utilization for POI Recommendations}\label{attribute}
	\end{figure}

	To investigate the varied information types employed as inputs in RSs, this study extends the taxonomy by \cite{9693280} and classifies pertinent information types during the tourism process into four principal data dimensions: Tourist, POI, Interaction and Context. By analyzing the collected papers, our study categorizes the utilized information types into these four data dimensions and provides a summary of their respective data elements. The schematic representation of this classification is presented in Figure~\ref{up_buttom}. To illustrate the frequency of usage of different information types, Figure~\ref{attribute} presents the proportions of papers that utilized at least one of the information types from the data elements listed in Figure~\ref{up_buttom}. Given the extensive collection of papers, this study highlights representative studies to illustrate more details about different information types used in POI recommendation, as shown in Table~\ref{statistics_info_type}.
	
	\begin{table}[p]
		\centering
		\caption{Detailed Utilization of Information Types in POI Recommendation}
		\renewcommand{\arraystretch}{1.2}
		\label{statistics_info_type}
		\setlength{\tabcolsep}{4pt}
		\rowcolors{4}{gray!25}{white}
		
		\begin{tabular}{|c|c|c|c|c|c|c|c|c|m{2.7cm}|}
			\hline
			\multicolumn{2}{|c|}{\textbf{Details}} & \multicolumn{7}{c|}{\textbf{Information Type}} &                                                                                                                                                                                                                                                                                                                                                                                                                                                                                                                                                                     \\
			\cline{1-9}
			\multirow{-5}{*}{Reference}            & \multirow{-5}{*}{\begin{tabular}{@{}c@{}}Model\\Acronym\end{tabular}}                     & {\rotatebox[origin=l]{90}{Check-in}}      & {\rotatebox[origin=l]{90}{POI Profiles}} & {\rotatebox[origin=l]{90}{\begin{tabular}{@{}l}Social Relationship\end{tabular}}} & {\rotatebox[origin=l]{90}{Feedback}} & {\rotatebox[origin=l]{90}{\begin{tabular}{@{}l}POI Visual Content\end{tabular}}} & {\rotatebox[origin=l]{90}{\begin{tabular}{@{}l@{}}Weather Context\end{tabular}}} & {\rotatebox[origin=l]{90}{\begin{tabular}{@{}l@{}}POI Popularity\end{tabular}}} & \multicolumn{1}{c|}{\multirow{-6}{*}{\textbf{\begin{tabular}{c}Input \\ Attribute\footnotemark[1]\end{tabular}}}} \\ \hline
			
			\cite{25-m}                                      & AGRAN                                          & $\surd $                                               &                                                            &                                                                                                          &                                                        &                                                                                                         &                                                                                                      &                                                                                                     & Geo. coordinates, \newline Timestamp                                 \\
			\hline
			\cite{28-m}                                      & QEXP                                           & $\surd $                                               & $\surd $                                                   &                                                                                                          &                                                        &                                                                                                         &                                                                                                      &                                                                                                     & Geo. coordinates, \newline Timestamp, \newline POI category                   \\
			\hline
			\cite{18-m}                                      & DPSR                                           & $\surd $                                               & $\surd $                                                   &                                                                                                          &                                                        &                                                                                                         & $\surd $                                                                                             &                                                                                                     & Geo. coordinates, \newline Timestamp, \newline POI category, \newline Weather condition \\
			\hline
			\cite{29-m}                                      & MAC                                            & $\surd $                                               & $\surd $                                                   & $\surd $                                                                                                 &                                                        &                                                                                                         &                                                                                                      &                                                                                                     & Geo. coordinates, \newline Timestamp, \newline POI category, \newline Friendship         \\
			\hline
			\cite{17-m}                                      & MEAL                                           & $\surd $                                               &                                                            &                                                                                                          &                                                        & $\surd $                                                                                                &                                                                                                      &                                                                                                     & Interaction matrix, \newline POI photo                                      \\
			\hline
			\cite{19-m}                                      & TeSP-TMF                                       & $\surd $                                               &                                                            &                                                                                                          &                                                        &                                                                                                         &                                                                                                      & $\surd $                                                                                            & Geo. coordinates, \newline Timestamp, \newline Number of visits       \\
			\hline
			\cite{30-m}                                     & Mandari                                        & $\surd $                                               &                                                            &                                                                                                          & $\surd $                                               & $\surd $                                                                                                &                                                                                                      &                                                                                                     & Geo. coordinates, \newline Timestamp, \newline POI photo, \newline Texual review        \\
			\hline
			\cite{7-m}                                      & FedPOIRec                                      & $\surd $                                               &                                                            & $\surd $                                                                                                 &                                                        &                                                                                                         &                                                                                                      &                                                                                                     & Interaction matrix, \newline Friendship                                      \\
			\hline
			\cite{12-m}                                      & DCDQN                                          & $\surd $                                               &                                                            &                                                                                                          & $\surd $                                               &                                                                                                         & $\surd $                                                                                             &                                                                                                     & Timestamp, \newline Texual review, \newline Weather condition                         \\
			\hline
			
			\cite{104-m}                                      & TransMKR                          & $\surd $                                               & $\surd $                                                           &                                                                                                          &                                         &                                                                                                         &                                                                                           &$\surd$                                                                                                       & Geo. coordinates, \newline Timestamp, \newline Number of visits and visiting users, \newline POI category                         \\
			\hline
			
			\cite{88-m}                                      & WUPSame                                          &                                               &   $\surd $                                                          &                                                                                                          & $\surd $                                               &                                                                                                         &                                                                                             &                                                                                                     & Rating score, \newline POI category                         \\
			\hline
			
			\cite{72-m}                                      & NATP                                          &                                             &                                                            &                                                                                                          & $\surd $                                             &                                                                                                         &                                                                                            &                                                                                                     & Texual review                      \\
			\hline
			
			\cite{64-m}                                      & CAFOB                                          &                                         & $\surd $                                                                  &                                                                                                          & $\surd $                                               &                                                                                                         &                                                                                            & $\surd$                                                                                                      & Geo. coordinates, \newline POI category, \newline Weather condition                         \\
			\hline
		\end{tabular}
		\footnotetext{Note: The $\surd$ symbol denotes that the proposed model utilizes the information type indicated in each column.}
		\footnotetext[1]{The column does not list the user and POI identifiers, which are required for all POI RSs. }
		
	\end{table}
	
	Upon analyzing the utilization of information types in POI recommendation research, a noteworthy tendency towards a limited amount of information types is presented. Over 50\,\% of the surveyed papers leverage check-in data, incorporating both timestamp and geographic information, as primary input data for their POI RSs. POI profiles (e.g., category and description of POIs) are the second most commonly utilized information type. Social relationship information, including friendship networks and social connections, are also utilized particularly in recent years, where graph-based approaches have gained prominence \citep{12,50,61}. Besides, user feedback (e.g., ratings and reviews) and other information (e.g., POI visual content, weather context and POI popularity) constitute smaller proportions of the utilized information types.
	
	All mentioned information types possess the potential to effectively capture tourists' preferences for POI recommendations. Collectively, these information can be classified into two categories: implicitly or explicitly provided information. Explicitly provided information refers to the data that directly reveals tourists' preferences, such as ratings, reviews and feedback comments. This type of information provides a deliberate, unambiguous and intentional quality assessment of user preferences, enabling the generation of recommendations that align with those preferences \citep{13470}. Implicitly provided information is typically inferred from user behavior and interaction patterns, such as clickstream data, search queries and consumption histories, which can provide valuable insights into tourists' preferences and interests as well \citep{Jannach2018}.
	
\begin{figure}[!htb]%
		\centering
		\begin{subfigure}[b]{0.49\textwidth}
			\centering
			\includegraphics[width=0.9\textwidth]{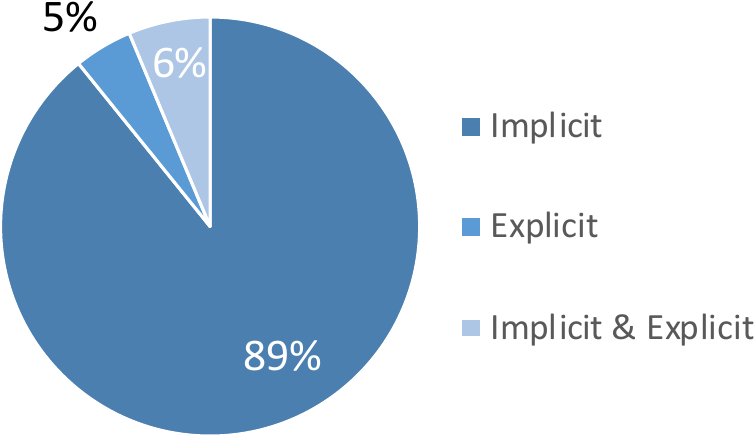}
			\caption{Explicit and Implicit Information }
			\label{feedback}
		\end{subfigure}
		\hfill
		\begin{subfigure}[b]{0.49\textwidth}
			\centering
			\includegraphics[width=1\textwidth]{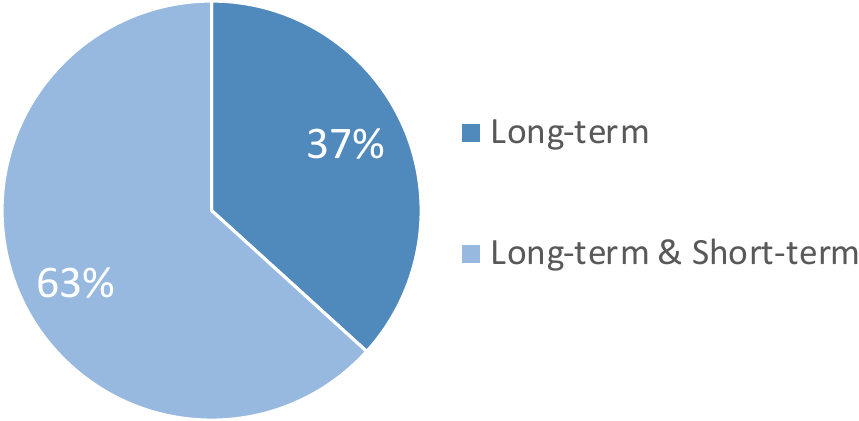}
			\caption{Long-term and Short-term Preferences}
			\label{term}
		\end{subfigure}
		\caption{Analysis of Information Feedback Types and Preference Terms in POI Recommendation Research}
	\end{figure}

	Explicitly and implicitly provided information offer distinct levels of expressivity regarding the user's preferences, and a combination of the two can lead to more accurate and effective recommendations \citep{1869453}. However, in the surveyed papers, as shown in Figure~\ref{feedback}, the majority of studies solely relied on a single type of information, either implicitly or explicitly provided information. Only a small percentage of the papers simultaneously utilized both types of information, such as incorporating check-in data along with reviews or ratings \citep{38,65,PANG2020106536}.
	
	An orthogonal question in the context of information on user preferences lies in their temporal dimension, i.e., long-term and short-term preferences. Long-term preferences are inherent and relatively stable, such as preferred activities and travel mode, which are influenced by the user's personal background, such as age, gender, education and income \citep{2348312}. Short-term preferences convey the user's tourism intention in a relatively short period and can be affected by transient events, such as impromptu short weekend trips or special personal occasions, like business travel. These preferences change more frequently and strongly compared to long-term preferences \citep{3295822}.
	
	Both long-term and short-term preferences contribute to providing precise and dynamic  POI recommendations. In this survey, long-term preferences are defined as the preferences that are inferred from a user's entire historical data, without making any distinctions based on time. In contrast, short-term preferences are defined as preferences derived by historical data with varying weights, where more recent data are given a higher priority. For example, a user's historical data might be divided into different sessions, with more recent sessions given higher influence weights in the prediction. Due to the widespread application of recurrent neural networks (RNN) across various domains, the traditional reliance solely on tourists' historical data to represent their long-term preferences for generating POI recommendations has been reduced. The analysis of the collected studies revealed that the majority of current research concurrently considers both long-term and short-term preferences of tourists (cf. Figure~\ref{term}). RNN and its variant such as Long Short-Term Memory (LSTM) and Gated Recurrent Units (GRU), have played a crucial role in this progress. These models are capable of handling sequential data, allowing them to leverage user's historical POI visit sequences. By inputting a user's check-in sequence into the model, it can learn patterns of user's temporal and spatial preferences \citep{11-m}. In addition, the attention mechanism\footnote{Attention mechanisms are rooted in the human visual attention system and were initially proposed in the field of visual image processing. The concept gained prominence with the publication by \cite{attention}, where self-attention mechanisms were extensively employed to learn textual representations.} also contributes significantly to this advancement because of its ability to enable the model to weigh different parts of the input sequence differently. This capability allows the model to focus on more relevant information, effectively capturing long-range dependencies and enabling the model to dynamically adapt to changing user preferences and contexts \citep{31-m}. It is likely the reasons why the short-term preferences of tourists have gained increased attention in recent years.
	
	\subsection{Techniques and Approaches in POI Recommendations}
	Given that research in POI recommendations predominantly relies on a handful of information types, it is equally worthy exploring how these data are utilized for POI recommendations. Building on the fundamental paradigms of recommendation systems introduced in Section \ref{Types of Recommender Systems}, this section analyzes current techniques and approaches in POI recommendation studies. The analysis aims to provide future researchers with an overview regarding the types of RSs used in POI recommendations, involved techniques and the reproducibility of recommendation models.
	
	\subsubsection{Recommender System Types}
\begin{figure}[!htb]%
		\centering
		\includegraphics[width=0.8\textwidth]{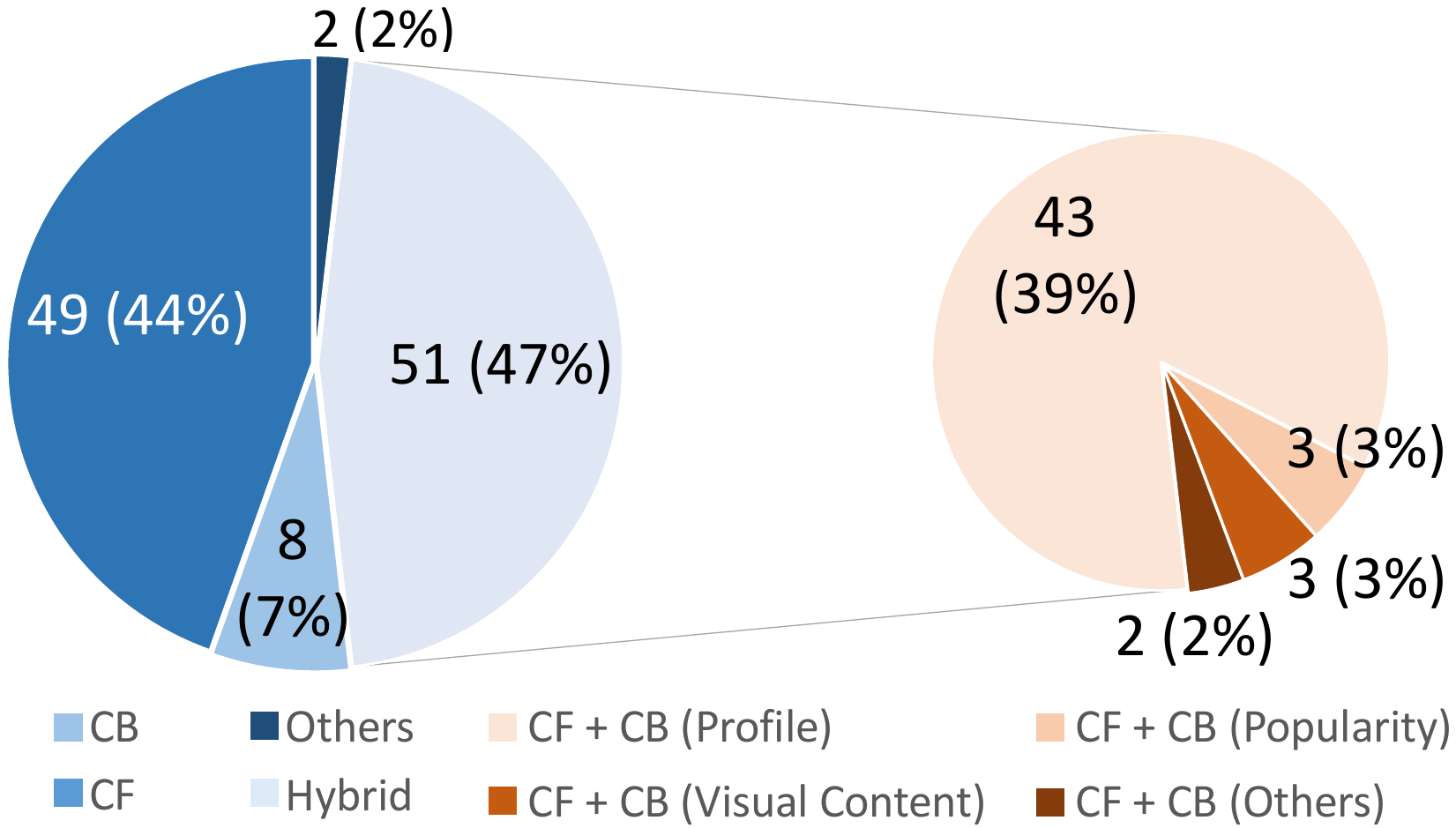}
		\caption{Techniques Employed in POI Recommender Systems}
		\label{rs_type}
	\end{figure}

	When we focus on the types of RSs employed in the collected POI recommendation research papers, it becomes evident that all three major families of recommendation approaches (CF, CB and hybrid techniques) are in use.
	The results of our analysis regarding the underlying technical approaches are shown in Figure~\ref{rs_type}.
	As the main approach in POI recommendations, CF are utilized in approximately 44\,\% of the research papers. Another substantial portion (47\,\%) of the research falls into the category Hybrid, which combines CF with additional POI contents beyond geographical information. This additional contents may encompass POI profiles (including category or description) \citep{59-m}, POI visual content \citep{17-m} or POI popularity information \citep{104-m}. Additionally, there are instances of pure CB POI recommendation research that construct POI profiles based on feedback about POIs \citep{15-m} or the categories of POIs \citep{65}, and subsequently provide recommendations of similar POIs based on these profiles. However, such studies constitute only a minority among the papers we collected. Finally, a few studies that do not align with the aforementioned taxonomy, such as knowledge-based RS \citep{KBRS} or conversational RS \citep{CRS}, are categorized as Others.
	
	It is worth noting that, as per definition provided in Section \ref{Types of Recommender Systems}, pure CF RSs exclusively rely on user-item interactions without any POI characteristics. However, the most frequently utilized user's historical check-in data in the process of POI recommendation often contains the geographical information about POIs. This content is frequently harnessed for tasks such as calculating the geographical distance between POIs \citep{27-m}. In this analysis, we classify such studies under the category CF, even when the research simultaneously utilizes geographical information about POIs. According to the collected papers, it is evident that recent research is increasingly focusing on harnessing a wide range of side information, including geographical information. Studies like those highlighted in \cite{26-m,23-m,10-m,14-m}, which do not depend on any geographical information and solely utilize check-in sequences, are relatively scarce.
	
	Among the categories discussed above, CF stands out as the most extensively utilized approach, demonstrating its prominence in the field.
	A possible reason is that the advent of neural networks has sparked considerable interest among researchers in recent years. Neural network architectures can be leveraged to train prediction models using the user-item interaction matrix, facilitating the generation of personalized recommendations. Consequently, the adoption of model-based CF has experienced a substantial upsurge, with the majority of CF studies incorporating such techniques. More detailed statistics will be presented in next Section~\ref{Recommendation Techniques and Approaches}.
	
	\subsubsection{Recommendation Techniques and Approaches}\label{Recommendation Techniques and Approaches}
	
	Upon a more granular examination of the methodologies employed in the collected POI recommendation studies, we classified them based on the taxonomy from \cite{3510409}. It is worth noting that some papers can be categorized into more than one category due to the simultaneous use of multiple methods. The detailed descriptions of these methods are provided as follows:
	
	\begin{itemize}
		\item Similarity-based methods: These techniques primarily rely on similarity functions, such as cosine similarity or Pearson correlation, to calculate similarity measures between users, items or both, in order to generate recommendations.
		
		\item Factorization-based methods: These approaches involve the factorization of user-item interaction matrices or auxiliary information-embedded tensors. Factorization-based methods aim to uncover latent factors that encapsulate user-item interactions and use these patterns for making recommendations.
		
		\item Probabilistic methods: These methods consider multiple random variables that may be related based on, e.g., Naïve Bayes, Kernel Density Estimation or other simple approximations\footnote{Typically these models rely on strong (and probably overly optimistic) independence assumptions \citep{3510409}}. In the context of recommendation, these random variables usually include users, items, and potential interactions between them.
		
		\item DL-based methods: These approaches utilize deep neural networks to capture intricate patterns in user behavior and item characteristics. These patterns are then used as the basis for generating recommendations.
		
		\item Graph-based methods: These techniques utilize network structures to model relationships between users and items. Users, items and their interactions are often represented as nodes and edges in a graph and recommendations are generated based on insights derived from graph-based representations.
		
		\item Other methods: This category encompasses research that employs recommendation techniques not falling into the above-defined categories, e.g., utilizing optimization techniques for POI recommendations \citep{66,16-m}.
	\end{itemize}
	
\begin{figure}[!htb]%
		\centering
		\includegraphics[width=0.8\textwidth]{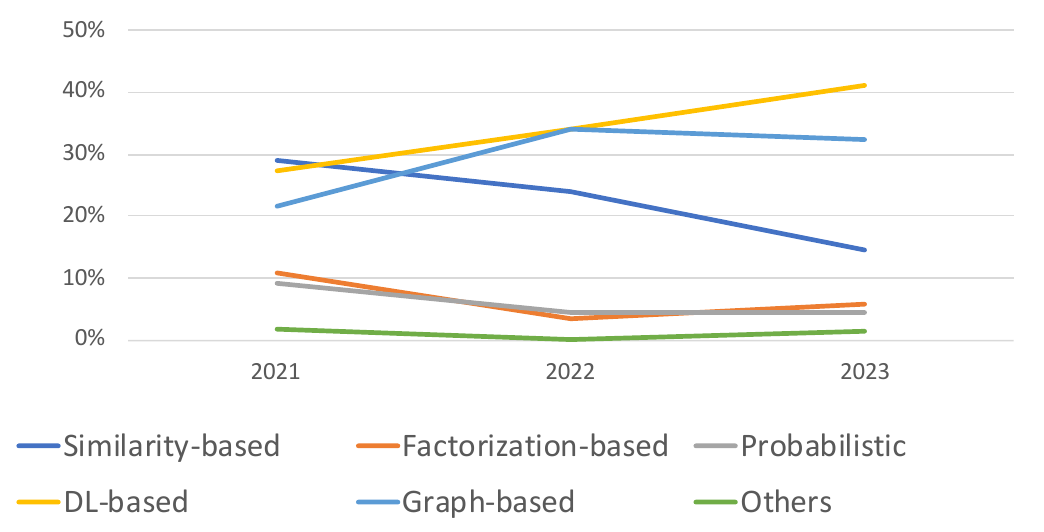}
		\caption{Trends in Utilization of Recommendation Methods in POI RSs}
		\label{technique}
	\end{figure}

	To illustrate the temporal evolution in the proportion of papers employing distinct methodologies, Figure~\ref{technique} presents the trend in RS research based on the aforementioned methods. In recent years covered by our study, similarity-based, DL-based and Graph-based methods have consistently constituted the majority compared to other methods. The core idea behind predominant research involves utilizing the user-item interaction matrix or the user's check-in sequence as input and leveraging convolutional neural networks (CNN) \citep{51}, RNN \citep{1} or graph neural networks (GNN) \citep{3-m} to learn representations from tourists and POIs. Subsequently, the similarity between POIs is calculated to offer a top-$n$ POI list as recommendations. Furthermore, some studies integrate similarity as part of the loss function, accounting for factors like geographical proximity between POIs \citep{22-m}, user group clustering \citep{75-m} or other embeddings\footnote{Embeddings refers to the mapping of high-dimensional objects (e.g., words, users, items, etc.) into a low-dimensional vector space. This mapping helps to capture semantic relationships and patterns between objects while reducing the dimensionality of the data, making it more feasible to compute and analyze.} \citep{2-m,10-m,42-m} in POI recommendations.
	Since similarity-based methods provide interpretable recommendations, these methods have contributed to a widespread application in POI recommendation.
	
	However, recent research suggests a gradual shift away from explicitly defining the similarity between POIs, as evidenced by a decreasing proportion of papers employing similarity-based methods. Instead, researchers are increasingly exploring the embeddings learned from deep neural networks or graph structures to directly generate POI recommendations among candidates. DL-based and graph-based methods exhibit the ability to capture complex relationships and associations among tourists and POIs, e.g., the research utilizing graph convolutional networks and attention mechanism \citep{25-m}. This capability facilitates the generation of more comprehensive and context-aware recommendations \citep{23,29,61}. The growing interest in exploring embedding techniques reflects an increasing attention for leveraging inherent data structures and dynamics to craft effective recommendations.
	
	Besides, some studies have employed probabilistic methods, such as utilizing kernel density estimation to estimate the distribution of tourists' interest features for POI recommendations \citep{18,63} and Factorization-based methods, e.g., tensor factorization incorporating contextual factors \citep{14}. However, these methods remain in the minority.
	
	While POI recommendation research based on DL has become dominant in the literature, and substantial progress over the state-of-the-art is claimed for all of them, it is worth noting that the computationally expensive neural methods do not always outperform well-established factorization-based methods or linear models \citep{DL-efficiency}. Since evaluating the effectiveness of each POI RS model is not the  focus of our work, readers can refer to \cite{DL-efficiency} and \cite{ferraridacrema2020tois} for more detailed insights.
	
	\subsubsection{Reproducibility of Recommendation Models}\label{Reproducibility and Progress}
	Available source codes encourage other researchers to replicate, verify and extend existing work, enhancing the credibility and reproducibility of research outcomes. Furthermore, in the context of applying modern DL techniques to recommendation problems, it was observed that the excitement for DL may have led to some rushed evaluations and a partially limited level of reproducibility, which may impact the achievement of true progress to a certain extent \citep{ferraridacrema2020tois}.  To see if similar problems may appear in POI recommendations, we pay particular attention to reproducibility aspects when reviewing the technical approaches.
	
\begin{figure}[!htb]%
		\centering
		\begin{subfigure}[b]{0.49\textwidth}
			\centering
			\includegraphics[width=\textwidth]{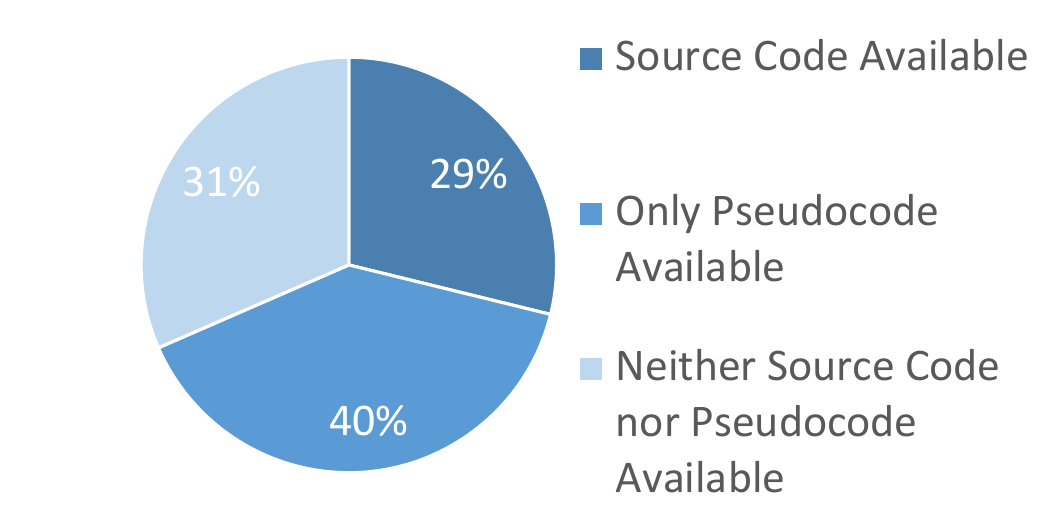}
			\caption{Code Availability in Collected Papers}
			\label{code_availibility}
		\end{subfigure}
		\hfill
		\begin{subfigure}[b]{0.49\textwidth}
			\centering
			\includegraphics[width=\textwidth]{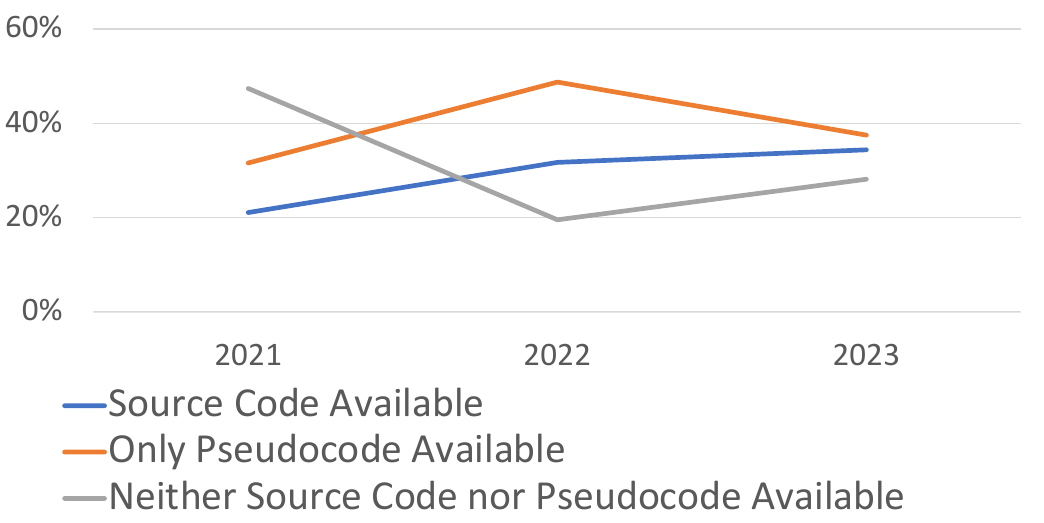}
			\caption{Trends in Code Availability}
			\label{code_availibility_year}
		\end{subfigure}
		\caption{Code Availability in POI Recommendation Research Papers}
	\end{figure}

	The availability of source code in the surveyed studies is depicted in Figure~\ref{code_availibility}. A notable observation is that 40\,\% of the papers solely provided pseudocode, which presents a high-level representation of the algorithms or methodologies without offering the actual implementation details; 31\,\% of the papers include neither the source code  nor pseudocode, thus restricting the ability to replicate and validate the research findings. Conversely, only 29\,\% of the papers provided the source code, which contributes to the reproducibility and transparency of the research process.
	In Figure~\ref{code_availibility_year}, we can observe that the proportion of papers providing source code has been on the rise. Nevertheless, the low level of code availability in the domain of POI recommendation research reflects the relatively low level of reproducibility.

	\subsection{Evaluation Approaches and Metrics  in POI Recommendations}\label{Evaluation Metrics and Approaches in POI Recommendations}
	When faced with a list of candidate POIs provided by the recommendation model, evaluating the quality of these recommendations poses a significant challenge. Evaluation approaches for RSs can be categorized into offline evaluations, lab/user studies and online A/B testing \citep{evaluation_dietmar}. Offline evaluations assess the performance of recommendation models through the user‘s historical data; lab/user studies typically encompass the invitation of a selected group of users to participate in an experiment, with subsequent collection and analysis of their feedback regarding the recommendations; online A/B testing entails randomly assigning users of a fielded system to groups using different recommendation algorithms and then comparing the user behaviors across these groups, such as click-through rates, purchase rates and user satisfaction levels to discern the effectiveness of the respective recommendation algorithms \citep{offlineE}. Next, we present the details of the different evaluation approaches utilized in POI recommendation research.
	
	\subsubsection{Offline Evaluation}
	Analyzing the collected papers, we found that the majority of research in the POI recommendation domain relies on offline evaluation.
	Offline evaluation, in contrast to the other two evaluation methods, focuses on evaluating the effectiveness of recommendation algorithms using pre-collected historical data. Although this approach assumes that ratings or implicit feedback in the held-out test set accurately define user preferences, it offers a quick and cost-effective means of evaluating the performance of a recommendation system \citep{3464733}. Thus, offline evaluation remains essential for investigating specific aspects of recommendation algorithms \citep{JannachPRZ21}.
	
\begin{figure}[!htb]%
		\centering
		\includegraphics[width=0.7\textwidth]{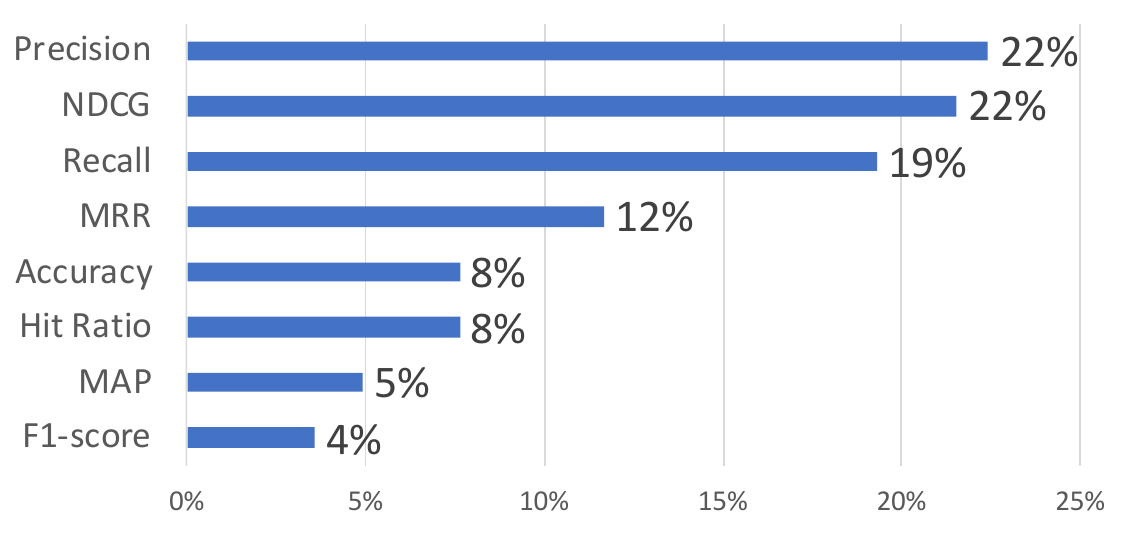}
		\caption{Evaluation Metrics for Relevance Measures in POI Recommendation Research}\label{evaluation}
	\end{figure}

	\paragraph{Evaluation Metrics}
	Offline evaluations enable the assessment of the recommendation quality from various perspectives, such as Relevance, Novelty, Serendipity, Diversity and Coverage \citep{RSTextbook}.\footnote{It is worth noting that, besides this evaluation perspective, there are other angles to assess the model's performance, such as Fairness \citep{RAHMANI2022117700}. Readers are referred to the relevant literature to obtain more details on this aspect \citep{fairness}.}
	\bmhead{Relevance}
	Relevance is an aspect to measure the ability of an RS to provide users with items that they find interesting \citep{RSTextbook}. It is essential to highlight that
	the relevance of recommendations is the predominant aspect when evaluating RSs \citep{beyondAcc}. The frequently employed evaluation metrics (utilized by more than five papers) for measuring relevance in the gathered POI recommendation studies are depicted in Figure~\ref{evaluation}. They can be classified into error-based metrics, (e.g., Precision, Recall and Accuracy) and ranking-based metrics (e.g., Normalized Discounted Cumulative Gain (NDCG), Mean Reciprocal Rank (MRR) and Mean Average Precision (MAP)) \citep{3510409}. According to the observed trends, the current evaluation of POI recommendation models extends beyond checking if recommended POIs are visited (error-based metrics). Ranking-based metrics are also widely used to assess the alignment between recommended and actual POI order in test sets. Upon comparing papers centered on general POI recommendation with those focused on next-POI recommendation, no notable differences are identified in terms of dominant metrics. Precision, Recall and NDCG underscore their prominent role in evaluating the relevance of recommendations.
	
	\bmhead{Novelty}
	Novelty is an evaluation aspect that measures how well a RS can recommend items that the users are not aware of, or that they have not seen before \citep{RSTextbook}. Especially in the tourism domain, recommending well-known POIs has limited value because even simple algorithms can point tourists to visit the most popular attractions, which are likely already known to them \citep{60}. A recommendation list containing highly popular POIs may lack substantial value for tourists.
	The most straightforward metric to evaluate the novelty of a RS is the average popularity of recommended POIs, under the assumption that less popular items contribute to higher novelty \citep{LU20121}.
	This is reflected in studies, where novelty is considered by analyzing the visit frequency of recommended POIs to determine their popularity  \citep{60, 28-m,62,44}.
	
	\bmhead{Serendipity}
	Serendipity in RSs is defined as the capability to suggest both relevant and unexpected items \citep{44-m}.
	Expanding on the unexpectedness measure, serendipity can be quantified by determining the proportion of recommendations that are both relevant and unexpected \citep{serendipity}.
	For instance, in the collected papers of our study, \cite{44-m} developed a serendipity-oriented model for the next-POI recommendation. This research measured the unexpectedness of recommendations through the similarity between POI categories and specific POIs, subsequently combining these measures with relevance metrics to assess the serendipity capability of the model.
	
	\bmhead{Diversity}
	A sophisticated RS aims to provide diverse recommendations to prevent user boredom with repeated and similar suggestions \citep{3527449}. Intra-user diversity, measuring diversity within a recommendation list for each user, is a widely explored evaluation metric \citep{LU20121}. In the context of the POI recommendation domain, \cite{14-m} assessed the model's diversity performance by evaluating the dissimilarity between different recommended POIs in the generated list. Consequently, diverse recommendations address the different preferences of users, ensuring a more comprehensive exploration of their interests.
	
	\bmhead{Coverage}
	Coverage serves as an important aspect to reflect whether the recommendation can cover the entire item space, user space or genre space \citep{diversity}. Higher coverage may benefit both tourists and business owners. Offering users a broader array of recommended items has the potential to enhance their satisfaction with the system \citep{coverage}, while concurrently boosting overall product sales \citep{coverageAdvan}.
	Among the collected papers, several studies, e.g., \cite{28-m,44,62}, considered coverage in the model evaluation stage. They examined the model's coverage from the perspective of the recommended POI space and the genre space based on POI categories.
	Although it is important to increase the coverage to improve the comprehensiveness of RS, the aforementioned POI recommendation studies also emphasized that it is important to reach a good balance between coverage and relevance.
	
	\paragraph{Evaluation Baselines}
	During the offline evaluation process of POI RSs, researchers often compare the performance of their systems against baseline approaches. In our analysis of the collected papers, we observed that a variety of baselines are used, indicating the absence of a universally accepted standard baseline for POI recommendation research. However, we identify several baselines that are chosen more frequently (more than 10 papers), which we summarize and present in Table~\ref{baseline}. The findings indicate a prevailing reliance on machine learning methods, as compared, e.g., to optimization-based methods, as chosen baselines in the evaluation of POI RSs. Additionally, due to potential challenges in terms of reproducibility, it is difficult to find a concrete implementation of a baseline algorithm from the state-of-the-art research in POI recommendations.
	
	\begin{table}[!htb]%
		\caption{Baseline Approaches for Evaluation of POI Recommender Systems}\label{baseline}
		\begin{tabular}{lp{10cm}}\toprule
			\textbf{Baseline} & \textbf{Description} \\  \midrule
			
			\multirow{2}{*}{FPMC} & A combination of common Markov chain and normal matrix factorization model \citep{FPMC} \\ \midrule
			\multirow{2}{*}{PRME} & A personalized ranking metric embedding method to model personalized check-in sequences \citep{PRME} \\ \midrule
			\multirow{2}{*}{LSTM} & A recurrent network architecture in conjunction with an appropriate gradient-based learning algorithm \citep{LSTM} \\ \midrule
			\multirow{2}{*}{ST-RNN} & Extended RNN that models local temporal and spatial contexts in each layer \citep{ST-RNN} \\ \midrule
			
			\multirow{2}{*}{STGN} & Extended LSTM adding spatial and temporal gates to capture user’s preference in space and time dimension \citep{STGN}  \\ \midrule
			\multirow{2}{*}{LSTPM} &  A model  that learns long term user preferences through a nonlocal network, and short-term user preferences with a geo-dilated network \citep{LSTPM} \\ \midrule
			\multirow{2}{*}{DeepMove} &  A model integrating a multi-modal embedding RNN and a historical attention mechanism \citep{DeepMove} \\ \midrule
			\multirow{2}{*}{STAN} &  A model using a bi-attention network for aggregating the spatial-temporal interval information in the check-in sequence \citep{STAN} \\
			\bottomrule
		\end{tabular}
	\end{table}
	
	\subsubsection{Lab/User Studies}
	Conducting lab/user studies provides a valuable advantage by capturing real-time user feedback and subjective preferences, offering insights into the user experience that cannot be fully addressed through offline evaluations alone. This approach enhances the understanding of how users interact with the recommendations in real-world scenarios \citep{RSTextbook}. Unfortunately, there has been limited exploration of lab/user studies in the field of POI recommendation research.
	Contrary to the widespread use of offline evaluations in POI recommendation research, only one study conducted by \cite{60} implemented lab/user studies to evaluate their POI RSs. The authors not only examined the model's offline performance but also designed an online user study to measure user-perceived novelty and appreciation of recommendations.
	
	\subsubsection{Online A/B Testing}
	Online A/B testing is a commonly used evaluation approach in the industry to assess the performance of a RS.
	Live experiments are conducted with control and treatment groups in a production environment. The primary objective is to compare the performance of two systems to make decisions based on their respective business values \citep{abtestOb}. In other RS domains, A/B testing has been researched \citep{businessValue}. Interestingly, in the field of POI recommendation, the application of online A/B testing is not observed in the papers identified in our work. This highlights the gap between POI recommendation research  and deploying a POI RS in production, which is frequently overlooked. The assessment of recommendation engines in real-world production settings frequently diverges from the offline evaluation. \citep{abTesting}. Therefore, more attention should be given to this evaluation aspect in future research.
	
	\section{Discussion}\label{Discussion}
	In this section, we further elaborate upon the results obtained in Section \ref{A Landscape of Research}, specifically discussing the application of multiple information types and methods for integrating heterogeneous data in the field of POI recommendation. We also identify existing gaps and explore potential research opportunities for tourism-related POI recommendation.
	
	\subsection{Utilization of Multiple Information Types in POI Recommendations}\label{The Role of Heterogeneous Data in in-trip POI Recommendations}
\begin{figure}[!htb]%
		\centering
		\begin{subfigure}[b]{0.49\textwidth}
			\centering
			\includegraphics[width=1.1\textwidth]{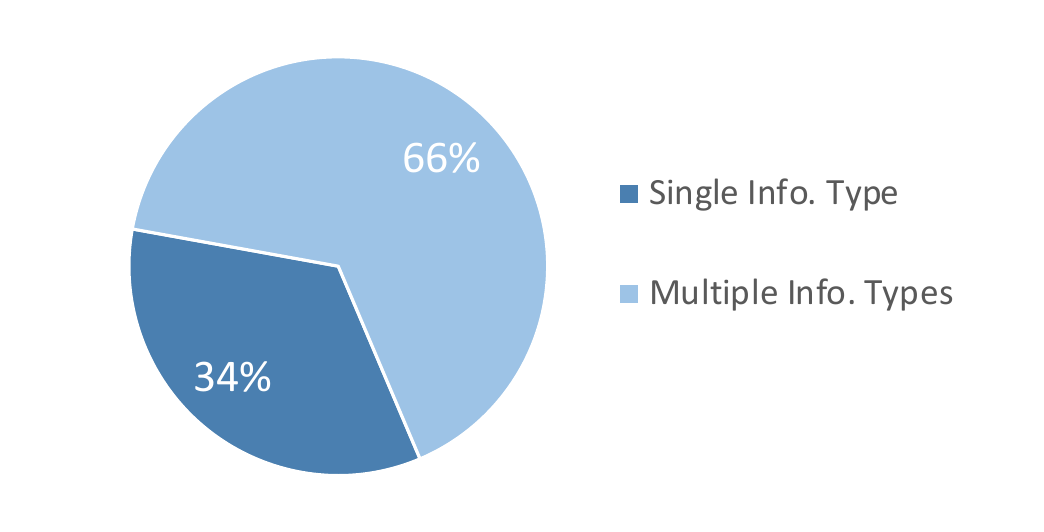}
			\caption{Proportion of Single versus Multiple Information Types}
			\label{heterogeneous}
		\end{subfigure}
		\hfill
		\begin{subfigure}[b]{0.49\textwidth}
			\centering
			\includegraphics[width=.9\textwidth]{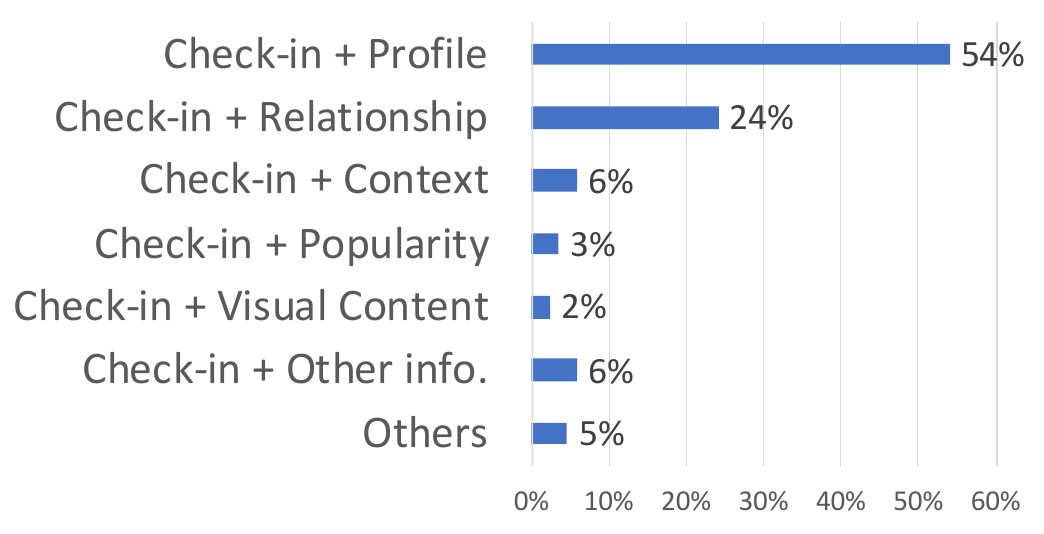}
			\caption{Distribution of Research Employing Multiple Information Types}
			\label{heatmap}
		\end{subfigure}
		\caption{Heterogeneous Data Utilized in POI Recommendation Research}
	\end{figure}

	In recent years, the availability of extensive and diverse information provides an opportunity for researchers to enhance the recommendation process by capturing different aspects of user preferences, POI characteristics and contextual information. However, looking at Figure~\ref{heterogeneous}, we observe that a substantial fraction of research, more than one third of the studied works, only rely on one single information type in the recommendation process. Most commonly, such approaches are based on historical user-item interaction data. As a result, these studies miss the opportunity to reach more accurate recommendations that may be obtained by considering other types of information. In contrast, the majority of studies already try to integrate multiple information types to decipher tourist preferences towards POIs. However, the diversity of information employed remains notably limited.
	Based on the analysis from the collected papers, we present the distribution of  information types utilized for POI recommendations in Figure \ref{heatmap}. The figure outlines the usage of different combinations of check-in data as primary information types with side information (such as POI profile, social relationship, etc.). Studies that do not leverage check-in data (for instance, those using reviews or ratings combined with other side information), are categorized as \emph{Others} in the diagram.
	
	It is noteworthy that as input data for POI RSs, the primary information types are check-in data and POI profiles (e.g., POI category \citep{21-m}), check-in data and tourists' social relationships (e.g., friendship graphs \citep{1-m}) or check-in data with both \citep{5-m}. The rest of the information types are less utilized (less than 5 of our collected papers). This tendency might be attributed to the limitation in publicly available datasets, which lack information related to tourists, POIs and contextual factors. Based on the presented analysis and relevant yet unmentioned aspects, the following challenges and shortcomings emerge in the field of POI recommendation especially when confronted with diverse information types.
	
	\bmhead{Insufficient consideration of biases in check-in data}
	According to Figure \ref{attribute}, check-in data is widely used as a primary information type in POI recommendation research and typically extracted from user activities on social networks \citep{checkIn}. This includes actions like checking in on LBSNs during POI visitation, providing reviews for visited POIs and capturing geo-tagged photos at POIs.
	Nevertheless, users may not necessarily disclose all activities that reflect their actual preferences on social networks, introducing a potential for inherent biases in check-in data \citep{Jannach2014AnalyzingTC}.
	Relying solely on check-in data while overlooking the biases within the data may potentially impact the real-world performance of POI RSs.
	
	\bmhead{Limited exploration of different information types}
	According to the analysis of the collected papers,  current research in POI recommendations does not fully exploit the potential of certain information types. For instance, visual content, textual reviews and contextual information are the less frequently utilized information types based on Figure~\ref{heatmap}.
	
	\begin{itemize}
		\item \textbf{Visual content:}
		Visual contents refer to visible information in photos captured by tourists during their in-trip phase. This may encompass landmarks, landscapes, people and other elements observed by tourists. The strong connection between visual contents and POIs can offer meaningful contextual information to implicitly deduce a tourists' preferences and needs, thereby enhancing the performance of POI RSs \citep{visualContent}.
		A mere four of the surveyed papers utilize POI visual content from public Flickr or Yelp datasets for POI recommendation research \citep{17-m,30-m,51,38}. These aforementioned studies share a common approach of utilizing extracted visual representations as a side information for POIs to improve the recommendation process. However, the visual content of POIs can offer much more, such as inferring user personality traits through user-uploaded photos \citep{VisualPersonality} or using all photos generated at a destination to construct projected and perceived destination images \citep{VisualCity}. Therefore, a deeper understanding of POI visual content provides significant potential for enhancing the performance of current POI RSs.
		
		\item \textbf{Textual reviews:}
		In the field of POI recommendation, relatively less attention has been given to individual user reviews. According to the papers collected in our study, only 9 out of 111 research studies attempted to incorporate reviews. The predominant approach to handling such data involves sentiment analysis on textual data extracted from user comments regarding POIs \citep{65}. However, compared to implicit feedback, such as check-in data, reviews provide explicit insights and offer a more accurate reflection of user preferences, especially when expressing dissatisfaction with specific items \citep{negative}.  The exploration of how to extract tourist's preferences from reviews remains a topic worthy of in-depth discussion.
		
		\item  \textbf{Contextual information:}
		Differing from the broad definition of contextual information by \cite{4}, contextual information in our paper specifically refers to environmental factors around POIs, such as weather information, traffic conditions and temporal information related to visiting times. The consideration of these contextual factors contributes to enhancing the personalization of recommendation systems by better adapting to tourists' current environments  and accurately capturing the dynamic changes in tourists' behaviors \citep{144}. However, among the collected papers in this survey, only a minimal number of studies (7 out of 111 papers) take contextual information into account. For instance, \cite{18-m} explored the impact of weather on user preferences and simulated the process of recommending POI sequences in a dynamic environment. However, the majority of POI research does not specifically address environmental factors associated with POIs. The reason may be that the datasets commonly used in POI recommendation research do not include the aforementioned contextual information.
	\end{itemize}

	\bmhead{Insufficient attention to popular travel platforms}
	According to statistics from Statista\footnote{\url{https://www.statista.com}} in 2023, the top five most popular travel and tourism websites worldwide (based on visit share) are booking.com\footnote{\url{https://www.booking.com}}, tripadvisor.com\footnote{\url{https://www.tripadvisor.com}}, airbnb.com\footnote{\url{https://www.airbnb.com}}, expedia.com\footnote{\url{https://www.expedia.com}} and agoda.com\footnote{\url{https://www.agoda.com}}.
	As indicated in Table \ref{platform}, the predominant data sources for POI recommendation research are LBSN platforms. Notably, the various data presented on the popular travel and tourism platforms above are not extensively covered. This gap may arise from the fact that these tourism platforms primarily concentrate on the pre-trip (hotel, flight) booking phase. Additionally, they do not offer publicly available  free APIs or share anonymized datasets for research, except in a few cases  (e.g., RecSys Challenge for hotel ranking\footnote{\url{https://recsys.acm.org/recsys19/challenge}}). However, this discrepancy should be brought to the attention of researchers to align POI recommendation studies with real-world application scenarios.
	
	\bmhead{Limited consideration of GDPR implications}
	In the domain of POI recommendation, careful consideration of the implications of the General Data Protection Regulation (GDPR)\footnote{\url{https://gdpr-info.eu}} is imperative, particularly given the involvement of sensitive user information, such as demographic information, usage history and sensitive geographical location \citep{8385136}. The GDPR imposes rigorous regulations on the processing of personal data to ensure comprehensive user privacy protection \citep{GDPR}. In the collection of papers examined in our study, only a few works have considered privacy protection issues during utilizing various data from tourists. For instance, in the work from \cite{9-m}, a data generation model was designed to replace user privacy-related data in the original dataset. \cite{75-m} explored the utilization of locality-sensitive hashing to anonymize POI information. Additionally, \cite{7-m} and \cite{21-m} employed a privacy-preserving federated learning framework, ensuring user's privacy while enabling effective POI recommendations. The primary objective of aforementioned research is to ensure that RSs make optimal use of tourists' information while aligning with privacy protection regulations. However, the challenge of ensuring user privacy in POI recommendation research seems to require more awareness and research.
	
	\subsection{Approaches to the Integration of Multiple Information Types in POI Recommendations}\label{Integration of Multiple Information Types in POI Recommendations}
	When confronted with diverse information types, the approaches employed by current POI recommendation research to integrate these heterogeneous data are worth exploring within our survey. From the research papers utilizing multiple information types, it is observed that the methods of integrating heterogeneous data vary according to the distinct recommendation approaches used in the respective studies.
	
	\subsubsection{Integration Through the Objective Function}
	For CF-based POI RSs, factorizing the user-item check-in matrix is a common method. To utilize information types beyond check-ins, a prevalent integration method involves constructing a unified objective function to incorporate the impact of heterogeneous data. For instance, \cite{18-m} extracted tourists' preferences and POI popularity with fine-grained contexts, enabling the model to capture the attractiveness of the POI sequence. The model considered user preferences and POI popularity in specific weather conditions and time periods, and incorporated them into the objective function of the POI recommendation model.
	
	A distinct advantage of this approach is that it can not only facilitate the integration of diverse information but also retain interpretability to a certain extent. However, a notable drawback is the potential limitation in capturing complex relationships among different information types. To address this, there is a growing necessity to explore more sophisticated integration methods.
	
	\subsubsection{Integration Through Embeddings}
	With the increasing adoption of DL-based methods in the field of POI recommendation, an increasing number of studies are leveraging advanced techniques to effectively integrate  diverse information types, capturing the intricate relationships and patterns inherent in the data.  For example, \cite{13-m} extracted contextual embeddings of users (mainly referring to social relations) and semantic embeddings of POIs, then fused them through a propagation process. The resulting representative vectors can serve as inputs for subsequent recommendation models.
	
	The utilization of embedding integration allows the capturing of complex relationships between user behaviors. This aids in a more comprehensive understanding and modeling of tourists' preferences in different contexts \citep{3}. Additionally, embedding integrates different information types into a common representation, enabling the model to automatically adapt to relationships between heterogeneous data without the need for manual feature engineering or rule design. For example, embedding-based universal user modeling approaches have been proven to be effective in integrating diverse information types \citep{Synerise}. However, employing embeddings for the integration of different data may involve computationally intensive processes, potentially resulting in relatively slow training and inference processes, especially with large-scale datasets \citep{embedding}.  In terms of interpretability, while DL-based models can capture intricate relationships, users and developers may find it challenging to understand how the models make specific recommendations.
	
	\subsubsection{Integration Through Graph-based Techniques}
	As depicted in Figure \ref{technique}, graph-based approaches  gain increasing interest in POI recommendation research. In a graph-based model, various information types are depicted as nodes, and the relationships between these information types are represented as edges. Subsequently, nodes and edges are mapped to a low-dimensional vector space, creating embedding representations as inputs for subsequent POI recommendations. This integration through graph structures enables flexible exploration of complex interdependencies among tourists, POIs and contextual factors. For instance, in a study conducted by \cite{50}, GNNs were employed to learn representations of tourists and POIs. In this study, each tourist is interconnected with others via social relations and with POIs via check-in activities. Subsequently, a latent representation of the target nodes was generated by merging the outputs of social neighbor integrations and POI neighbor integrations through a neural network.
	
	The application of graph-based integration provides a flexible framework for handling heterogeneous data in POI recommendation, offering a unified approach to learning representations. Moreover, it explicitly encodes the crucial collaborative signal of user-item interactions to enhance user/item representations through the propagation process \citep{graphIntegration}. However, compared to embedding integration approaches, graph models typically involve a large number of nodes and edges, resulting in higher computational complexity. Therefore, graph-based integration in the field of POI recommendation still faces challenges in terms of scalability and robustness \citep{graphIntegration}.
	
	\subsection{Advancing POI Recommendation Research}
	Based on the trends observed in the collected papers and the discussed challenges, we identify certain research gaps within the current body of work. Correspondingly, we propose potential research opportunities to drive advancements in the domain of POI recommendation.
	
	\subsubsection{Towards More Information-Rich Models}
	According to the results presented in Figure \ref{heterogeneous}, the majority of studies attempts to utilize multiple information types to understand tourists' preferences in current POI recommendation research. However, these studies still heavily rely on check-in data and there remains a lack of comprehensive exploration toward integrating multiple information types. Various information types could lead to significantly improved POI recommendations, especially when facing data sparsity problems. Leveraging side information is expected to address this problem. For instance, trying to simultaneously consider diverse aspects of tourists, POIs and context and leveraging the techniques discussed in Section \ref{Integration of Multiple Information Types in POI Recommendations} to integrate these heterogeneous data has the potential to enhance the accuracy and effectiveness of recommendations.
	
	The current state of this insufficient exploration of multiple information types is primarily attributed to the constraints posed by today's datasets. As shown in Table \ref{platform}, widely used datasets contain limited information types. For example, the Foursquare dataset only includes check-in timestamp, coordinates and POI category information. Therefore, exploring datasets with more diverse information types could be considered as a standard benchmark for future POI recommendation research, facilitating comparisons between different models.
	
	Since data from popular travel-related platforms did not appear
	in the current literature on POI recommendation, seeking collaboration with these platforms to introduce datasets rich in side information into academic research is worth exploring in the future. However, the reality is that many platforms refrain from providing User Generate Content (UGC) due to privacy concerns and economic reasons. It could however be highly beneficial if tourists have the option to share their published content publicly for research purposes (much like how Flickr users can choose different licenses for their uploaded photos \citep{2812802}), since the ownership of UGC should reside with the tourists rather than the platforms.
	
	\subsubsection{Enhancing the Reproducibility and Transparency of Models}
	As observed in Figure~\ref{code_availibility}, the availability of source code is limited in a significant number of papers.
	Open access to source code allows for the reproducibility, verification and extension of existing methodologies. The reproducibility not only strengthens the validity and credibility of research but also encourages collaboration and knowledge sharing among researchers.
	Therefore, it is critical for future research to place a stronger emphasis on ensuring the availability of source code and relevant documentation, to facilitate the reproducibility of published research works.
	
	As shown in Figure \ref{technique}, DL-based methods dominate in POI recommendation research. However, RSs using DL-based methods are complex black-box models, lacking interpretability and transparency of decision making \citep{XAI}. This poses a challenge for users and developers to understand how the models operate. Therefore, in addition to the mentioned ways of improving transparency through open datasets and source codes, it is necessary to develop more interpretable and understandable recommendation models, such as with help of Explainable Artificial Intelligence (XAI) techniques. Such models can help users and developers to better understand why specific recommendations are made, thereby increasing trust in the system.
	
	\subsubsection{Utilizing Diverse Evaluation Metrics and Approaches}
	In current POI recommendation research, there exists an overreliance on relevance measures in offline evaluation and it exhibits an underutilization of diverse evaluation metrics, as emphasized in Subsection \ref{Evaluation Metrics and Approaches in POI Recommendations}. For example, a general POI recommender system might recommend a first-time tourist to Paris to visit the Eiffel Tower, although the user most likely already knows about this popular attraction. Similarly, when making a next-POI recommendation, it might be predicted that the tourist will visit a nearby cafe based on his current location. While these recommendations might yield high accuracy in offline evaluations, the value of such recommendations to tourists can be arguably low. While relevance is of importance, it is equally imperative to recognize the multidimensional nature of the POI recommendation quality, along with the varying preferences and requirements of tourists. Hence, there is a pressing need for future research to diversify evaluation metrics, such as novelty, diversity, serendipity and coverage, to capture the true value and usefulness of recommendations in the specific context of tourism.
	
	Furthermore, offline evaluation primarily focuses on the model's performance on historical data and may not reflect the performance of the RSs in real-world scenarios. Indeed, offline evaluation may fail to capture user's subjective experiences, personalized needs and recommendation diversity. In contrast, lab/user studies can provide in-depth insights into user's expectations, satisfaction and usage experience \citep{offlineE}. This practical approach can compensate for the limitations of offline evaluation, making the RSs more aligned with real user needs. Therefore, in future POI recommendation research, researchers could incorporate more lab/user studies to evaluate the model's performance, while simultaneously executing offline evaluation.
	
	\section{Summary}\label{Conclusion}
	In this work, we conduct an analysis of the current state of research on POI recommendations. By addressing RQ1, we gain initial insights into the prevalent data utilization, techniques and evaluation methodologies in POI recommendation research. This provides an understanding of the latest research trends in this domain over the past three years. To tackle RQ2, we direct our attention to the utilization of multiple information types and integration approaches for heterogeneous data in the field of POI recommendation. This revealed an excessive reliance on check-in data, with limited exploration of other information types. Finally, based on the results derived from the analysis, we identify existing gaps and propose potential future directions from the perspectives of data, techniques and evaluation for RQ3.
	
	As the first information-centric survey on POI recommendation research, our paper serves as a reference for researchers, providing a foundation for the development of increasingly accurate, personalized and context-aware RSs. However, it is important to note that our study has some limitations. Due to the limited use of heterogeneous data in tourism-related POI recommendation research, this survey did not gather sufficient data to further explore the unique features of heterogeneous data in POI recommendation. Additionally, while our focus is on heterogeneous data used in POI recommender systems, the application of different recommendation algorithms for various information types has not been thoroughly investigated. And the impact of integrating different information types on the evaluation of POI RSs also remains an open question. These limitations highlight areas that warrant further exploration in future research.
	

\end{document}